\documentclass[12pt,preprint]{aastex}
\usepackage{amsmath,amssymb,graphicx}
\usepackage{epsfig}
\usepackage{graphicx}
\usepackage{color}
\usepackage{enumerate}
\usepackage{ulem}
\usepackage{epsfig}
\usepackage[] {placeins}

\def\lamrf{\lambda_{\rm rf}}
\def\Mbh{M_{\rm BH}}
\def\Redd{R_{\rm EDD}}
\def\Lbol{L_{\rm bol}}

\begin{document}
\title{ The correlations between optical variability and
physical parameters of quasars in SDSS Stripe 82}
\author{Wenwen Zuo, Xue-Bing Wu, Yi-Qing Liu, \& Cheng-Liang Jiao \\
Department of Astronomy, Peking University, Beijing 100871, China}

\begin{abstract}
   We investigate the optical variability of 7658 quasars from SDSS Stripe 82. 
Taking advantage of a larger sample and relatively more data points for each quasar, 
we estimate variability amplitudes and divide the sample into small bins of redshift, 
rest-frame wavelength, black hole mass, Eddington ratio and bolometric luminosity 
respectively, to investigate the relationships between variability and these parameters. 
An anti-correlation between variability and rest-frame wavelength is found. 
The variability amplitude of radio-quiet quasars shows almost no cosmological evolution, 
but that of radio-loud ones may weakly anti-correlate with redshift. In addition, 
variability increases as either luminosity or Eddington ratio decreases. 
However, the relationship between variability and black hole mass is uncertain; 
it is negative when the influence of Eddington ratio is excluded, but positive when 
the influence of luminosity is excluded. The intrinsic distribution of variability amplitudes 
for radio-loud and radio-quiet quasars are different. Both radio-loud and radio-quiet quasars 
exhibit a bluer-when-brighter chromatism. Assuming that quasar variability is caused by variations 
of accretion rate, the Shakura-Sunyaev disk model can reproduce the tendencies of observed correlations 
between variability and rest-frame wavelength, luminosity as well as Eddington ratio, supporting that 
changes of accretion rate plays an important role in producing the observed optical variability. 
However, the predicted positive correlation between variability and black hole mass seems to be 
inconsistent with the observed negative correlation between them in small bins of Eddington ratio, 
which suggests that other physical mechanisms may still need to be considered in modifying the simple 
accretion disk model.
\end{abstract}

\keywords{galaxies: active --- galaxies: nuclei --- quasars:
general --- quasars: variability}

\section{Introduction}
     Variability is one of the major characteristics of Active Galactic Nuclei
(AGNs), whose luminosities vary in all the bands from $\gamma$-ray to
radio, on timescales from hours to years. Although the study of variability
plays an important role in investigating the nature of the compact central
region in AGNs \citep{Hook94, Wilhite08}, the mechanism underlying quasar
variability is still inconclusive. Several physical mechanisms have been proposed
to explain quasar variability \citep{VB04}, including accretion disk
instabilities \citep{Rees84, Kawaguchi98, Kato96, Manmoto96, Czerny08},
Poissonian processes, such as multiple supernovae or star collisions
\citep{Terlevich92, Courvoisier96, Torricelli00}, and gravitational
microlensing \citep{Hawkins93}.

     To clarify the nature of quasar variability, most previous studies
focused on the dependencies of the variability indicator on redshift,
time lag, rest-frame wavelength and luminosity.
A decrease of variability was found towards longer wavelengths
\citep{Cutri85, Paltani94, Di96}, but the influences of other
parameters on the correlation were not excluded due to small samples.
\cite{VB04} confirmed such an anti-correlation by excluding the dependencies
of variability on redshift and luminosity, however, for the scarcity of data points
of each quasar, they estimated one variability indicator from a sample of quasars
(the ensemble variability method) rather than from an individual quasar (the
individual variability method). This ignores the diversity of an individual quasar
light curve, which may eventually fade out the actual relations. Due to the statistical
incompleteness of small samples, previous studies found controversial evidences for
the dependencies of variability on redshift and luminosity
\citep{Bonoli79, Netzer83, Cristiani90, Giallongo91, Hook94}.
Thus, the results may not be statistically reliable due to possible
dependencies of variability on other parameters. Only recently, with
larger samples, some work managed to disentangle
the anti-correlation between the variability amplitude and luminosity
by restricting other parameters in small ranges \citep{Wilhite08, Bauer09}.
However, they still adopted the ensemble variability method to estimate the variability
indicator. Moreover, as \cite{Wilhite08} only studied quasars with redshift larger than
1.69, the situation for quasars with lower redshifts were still unknown.
Besides luminosity, more attention has been paid to investigate the influences on variability
from other intrinsic parameters, i.e., black hole mass and Eddington ratio (the ratio of
bolometric luminosity to Eddington luminosity). \cite{Wold07} found a correlation between
variability and black hole mass from about 100 quasars, without accounting for the
influences of other parameters. Based on their ensemble variability study
of nearly 23,000 quasars from the Palomar QUEST Survey, \cite{Bauer09} found the dependencies
of quasar variability on black hole mass (positively), and Eddington ratio (negatively). Similar
conclusions were reached by \cite{Ai10} based on the individual variability indicators but without
disentangling the relations between these parameters.

   Although variability depends on many parameters, the underlying physical mechanisms
are still uncertain. Dividing the sample according to different parameter bins is essential
to analyze their influences on variability.
Moreover, to avoid missing the diversity of the individual quasar light curve,
the detailed variability indicator of each quasar rather than the ensemble variability
indicator from a group of quasars, should also be utilized.
However, no previous work has employed the two essential methods simultaneously in their
correlation analysis.

    In this paper we revisit the correlations between the variability amplitude
and physical parameters based on a sample of 7658 quasars in SDSS Stripe 82,
with both the individual variability method and the detailed parameter binning techniques.
We describe the sample selection and quasar variability estimations in \textsection~\ref{sample}.
The variability distribution and color-brightness relationship are given in \textsection~\ref{colorbrightness}.
\textsection~\ref{wavelength} shows the correlation between the variability amplitude
and rest-frame wavelength for radio-loud and radio-quiet quasars.
\textsection~\ref{redshift} shows the correlation between variability and
redshift. We disentangle the dependencies of variability on Eddington ratio, luminosity
and black hole mass in \textsection~\ref{param}. Implications of our results and
the comparison with the accretion disk model are presented in
\textsection~\ref{discussion}. We summarize our results in \textsection~\ref{conclusion}.
The cosmological parameters of ($h_0$, $\Omega_0$, $\lambda_0$) are set to be
(0.73, 0.27, 0.73) throughout the paper.

\section{Quasar sample and variability estimation\label{sample}}
\subsection{Sample}
A sample of 9254 variable quasars is obtained from cross-matching the spectroscopic
confirmed SDSS DR7 quasars \citep{Schneider10, Shen11} and a sample of 67507
variable sources in SDSS Stripe 82, which lies along the celestial equator
in the Southern Galactic Hemisphere (22h 24m~$< \alpha_{J2000} <$~04h 08m,
$-1.27\arcdeg<\delta_{J2000} < +1.27\arcdeg$, $\sim290$ deg$^2$)
and have repeated photometric observations (at least 4 per band, with
a median of 10) measured in up to 10 years in the $u^\prime g^\prime
r^\prime i^\prime z^\prime$ system \citep{Fukugita96, MacLeod10, Ivezic07, Sesar07}.
Black hole masses, bolometric luminosities and Eddington ratios are obtained from
the quasar catalog in \cite{Shen11}, where the black hole masses of quasars were settled by
fiducial virial mass estimates: H$\beta$ \citep{Vestergaard06} estimates for $z < 0.7$,
[Mg{\footnotesize II}] \citep{Shen10} estimates for $0.7 \le z < 1.9$ and
C{\footnotesize IV}] \citep{Vestergaard06} estimates for $z \ge 1.9$.
If more than one emission line are available, we still adopt the fiducial virial
black hole mass in \cite{Shen10}. In total, the quantity is measurable for
9148 quasars (98.9\%) in the sample.

\subsection{Variability estimation}
To measure the variability amplitude in each filter band for each quasar, we adopt the
formalism similar to that used in \cite{Ai10} and \cite{Sesar07}:
\begin{equation}
\Sigma = \sqrt{\frac{1}{n-1}\sum_{i=1}^{n}(m_{\rm i}-\langle m \rangle)^2}
\end{equation}

\begin{equation}\label{eq4}
V=\left\{
   \begin{array}{ll}
      (\Sigma^2-\xi^2)^{1/2}, & { if~\Sigma > \xi,}\\
      0, & { otherwise} .\\
   \end{array}
\right.
\end{equation}
where $n$ is the number of observations for a single source in each band, $m_{\rm i}$ is the
magnitude from the $i_{\rm th}$ observation,
$\langle m \rangle$ is the mean magnitude, and $\xi$ is estimated from
the photometric error $\Delta m_i$:
\begin{equation}
\xi^2=\frac{1}{n}\sum_{i=1}^{n}{(\Delta m_{\rm i})^2}
\end{equation}

We adopt $V$ calculated in this way as the variability amplitude, which describes the
intrinsic quasar variability by excluding the photometric error.
Before estimating the variability amplitude in one band for an individual quasar,
we eliminate the photometric outliers to make sure the calculated variability amplitude is not
caused by non-physical reasons \citep{Schmidt10, MacLeod10, Butler11}.
Assuming that a significant decrease in magnitude in a single observation may be non-physical
\citep{Schmidt10}, we exclude data points with very faint luminosity, i.e., magnitude fainter than 30.
Data points with large photometric uncertainties (errors larger than 1 magnitude) are also excluded.
Photometric measurements are marked as outliers if the residuals between them and their median 
magnitude of all the preliminarily cleaned data points are above 5 times their standard 
deviation. This procedure is iterated until all the outliers are excluded, each time removing only 
the strongest outlier and updating the median magnitude along with the standard deviation 
for the remaining measurements. After eliminating all the assumed
non-physical outliers, if more than 10 cleaned data points remained, we calculate the variability 
amplitude with the equations mentioned above. Otherwise, variability amplitudes are set to be 0.
To get more reliable statistical results, in the following studies we only investigate quasar
variability amplitudes in the SDSS $g^\prime$, $r^\prime$ and $i^\prime$ band, as photometric
errors in the $u^\prime$ and $z^\prime$ band are substantially larger \citep{Sesar07}.
Among the 9148 quasars with measurable black hole mass \citep{Shen10}, 8967 quasars have reliable
variability indicators in the $g^\prime$, $r^\prime$ and $i^\prime$ band, which means that variability
in each band for a quasar is a non-zero value estimated from at least 10 data points, even after
eliminating photometric outliers. 

\section{Variability distribution and color changes\label{colorbrightness}}
Since the physical origin of optical variability in radio-quiet and radio-loud quasars may
be different \citep{Giveon99}, among the 8967 quasars, we divide the 7658 quasars with either
radio detections or in the fields of the FIRST radio survey into two subsamples \citep{White97}:
the radio-quiet subsample contains quasars with radio loudness below 10; the radio-loud subsample holds
quasars with radio loudness above 10. Here radio loudness is defined as the ratio of
the observed radio flux density at rest-frame $6$ cm  and the optical flux density
at rest-frame 2500 \AA \citep{Becker95, Shen11}. 416 radio-loud sources and 7242 radio-quiet
sources are obtained for the study in the following sections.

\begin{figure}[]
\plotone{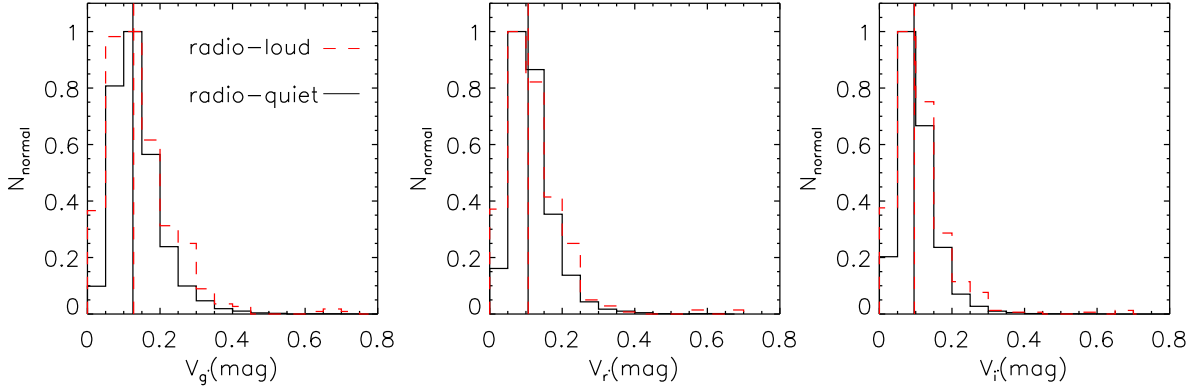}
\caption{Normalized distribution of the variability amplitudes in the $g^\prime$,
$r^\prime$ and $i^\prime$ band for the 416 radio-loud sources (red dashed lines)
and 7242 radio-quiet sources (black solid lines). The median values are shown with the
vertical lines.
\label{sigma_distribution}}
\end{figure}

     From Fig.~\ref{sigma_distribution}, we can see that the variability amplitudes
of the radio-quiet and the radio-loud quasars are mostly distributed between
0.05 mag and 0.3 mag, except that there is a relatively higher fraction of the radio-loud
quasars ($\sim$ 2.96\%) with variability amplitudes larger than 0.3 mag compared to
that of the radio-quiet quasars ($\sim$ 1.73\%).
The median variability amplitude of the radio-quiet objects in the $g^\prime$, $r^\prime$ and
$i^\prime$ band are 0.125$\pm$0.071 mag, 0.106$\pm$0.060 mag and 0.095$\pm$0.057 mag and that of the radio-loud
objects are 0.127$\pm$0.096 mag, 0.106$\pm$0.081 mag and 0.096$\pm$0.078 mag respectively.
Therefore, the median variability amplitude of the radio-quiet objects is very similar
compared to that of the radio-loud quasars. However, their distributions differ significantly according to
the results of 100 Kolmogorov-Smirnov (KS) tests (the median $p$ $\sim$ 0.012). 
Each time we randomly split both the radio-quiet and radio-loud quasar samples into half and 
perform KS tests on the generated samples. For variability values in the $g^\prime$, $r^\prime$ and 
$i^\prime$ band, 99.7, 91.7 and 81.5 percent of p values from these KS tests are less than 0.05 respectively, 
further indicating significant differences between their intrinsic distributions. Moreover, for both 
these radio-quiet and radio-loud quasars, the median variability amplitude decreases systematically 
as the observed wavelengths increase (see \textsection~\ref{wavelength} for more details).

\begin{figure}[]
\plotone{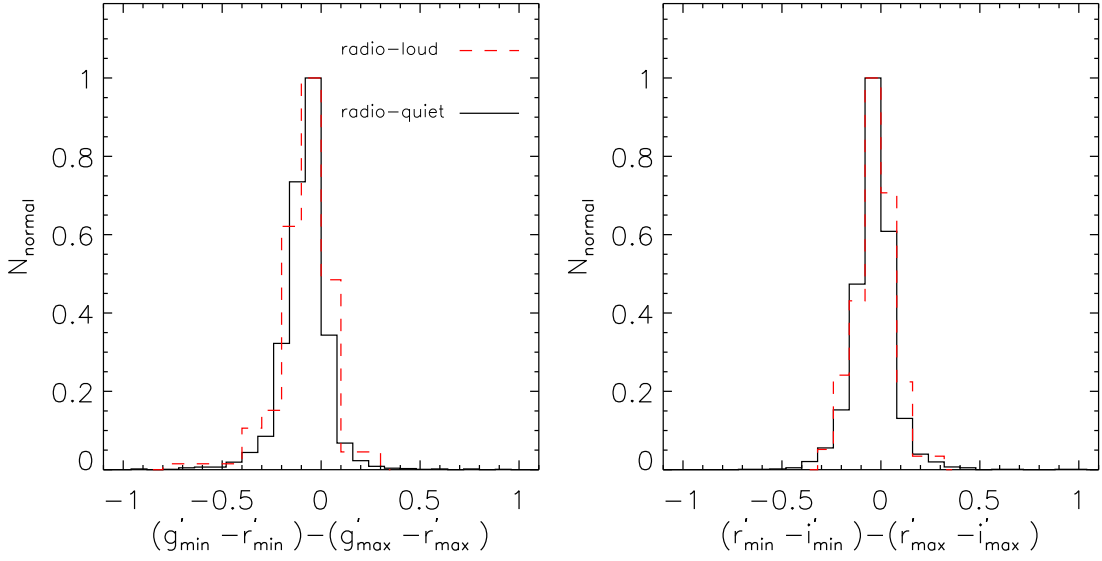}
\caption{Normalized distributions of the color differences between the brightest
state and faintest state. There are 166 radio-loud (red dashed lines) and 2811
radio-quiet quasars (black solid lines) in the left panel, 158 radio-loud and 2553 
radio-quiet quasars in the right panel, respectively.
\label{color_brightness}}
\end{figure}

     Here we take the magnitude difference between two adjacent SDSS filter bands
as the color indicator. Using the data in the $g^\prime$, $r^\prime$ and $i^\prime$ filter band,
we can get 2 colors, namely $g^\prime$-$r^\prime$ and $r^\prime$-$i^\prime$.
If both the difference of time in which the quasar luminosity of two adjacent bands show their
brightest state and the difference of time when luminosity of
two adjacent bands show their faintest state are less than 2 days, then these quasars are selected
to study the quasar color behavior. For the color $g^\prime$-$r^\prime$ and $r^\prime$-$i^\prime$,
we find 2811, 2553 radio-quiet quasars and 166, 158 radio-loud ones respectively. The differences
between colors in the brightest state and faintest state are shown in Fig.~\ref{color_brightness}.
Both the radio-quiet and radio-loud quasars display a strong bluer-when-brighter chromatism;
namely 82.9\%, 67.8\% of the radio-quiet quasars and 77.1\%, 63.3\% of the radio-loud quasars become
bluer in the $g^\prime$-$r^\prime$ and $r^\prime$-$i^\prime$ color respectively, when they become
brighter.

\section{Dependence on the rest-frame wavelength\label{wavelength}}
    As quasar variability amplitudes depend on quasar bolometric luminosity ($\Lbol$),
rest-frame wavelength ($\lamrf$), black hole mass ($\Mbh$) and
possibly redshift ($z$), such dependencies can be formulated as
$V$($\lamrf$, $z$, $\Mbh$, $\Lbol$),
where the combination of dependencies of the variability amplitude on
$\Mbh$ and $\Lbol$ can be replaced by the combination of its dependencies on $\Mbh$ and
$\Redd$ (Eddington ratio) or that on $\Redd$ and $\Lbol$. Dependencies of the variability amplitudes
on each parameter can be obtained in subsamples where other parameters are constrained
within small ranges. 
Some subsamples or sub-subsamples obtained from the combinations of those parameters yield
unphysical (imaginary) correlation results. In most cases when this occurs the number of quasars
is less than 10. In order to obtain statistically significant results, we require at least 10 quasars
in a data set before including it in the analysis. For all further analysis, binned data sets
including fewer than 10 quasars are rejected.
The bin size of $z$ is 0.3 and that of $\Redd$, $\Mbh$ and $\Lbol$ are all 0.5 dex. The minimum values 
of the four parameters are respectively 0.08, $10^{7.0}$ M$_{\odot}$, $10^{44.3} $erg$/$s and $10^{-3.13}$, 
while the maximum values are 5.08, $10^{10.8}$ M$_{\odot}$, $10^{48.0}$ erg$/$s and $10^{0.7}$. In this work, 
the ranges of the four parameters are chosen to be $[0.08, 4.9]$, $[10^{7.5}$ M$_{\odot}$, $10^{10.5}$ 
M$_{\odot}]$, [$10^{44.5}$ erg$/$s, $10^{48.0}$ erg$/$s] and [$10^{-2.0}$, $10^{0.5}$] respectively, as shown in 
Table~\ref{parambin}. More than 99.5 percent of quasars can be covered in this way and further 
expansion of the ranges does not yield more qualified subsamples.
Supposing that $X$ is one of the five parameters ($\lamrf$, $z$, $\Mbh$,
$\Lbol$ and $\Redd$), Spearman's Correlation Coefficient $r1\_X$
is calculated in each corresponding subsample or sub-subsample to describe the
rank correlation, while the significance $p\_X$ of its value deviating away
from 0 is also obtained. To parameterize the observed relation between
the variability amplitude and the parameter $X$, we fit all the data in each subsample
with a simple function:
\begin{equation}  \label{lamequation}
V(X) = b\_{X} \log{X} + a\_{X}
\end{equation}
Accompanied by the slope $b\_X$ and y axis intercept value $a\_X$, Pearson
Product-Moment Correlation Coefficient $r2\_\log{X}$ is also calculated to illustrate
the strength of the proposed relation ($r2\_X$ for short). In the following sections,
$X$ will be replaced with $\lamrf$, $z$, $\Mbh$, $\Lbol$ and $\Redd$.

\begin{figure}[]
\plotone{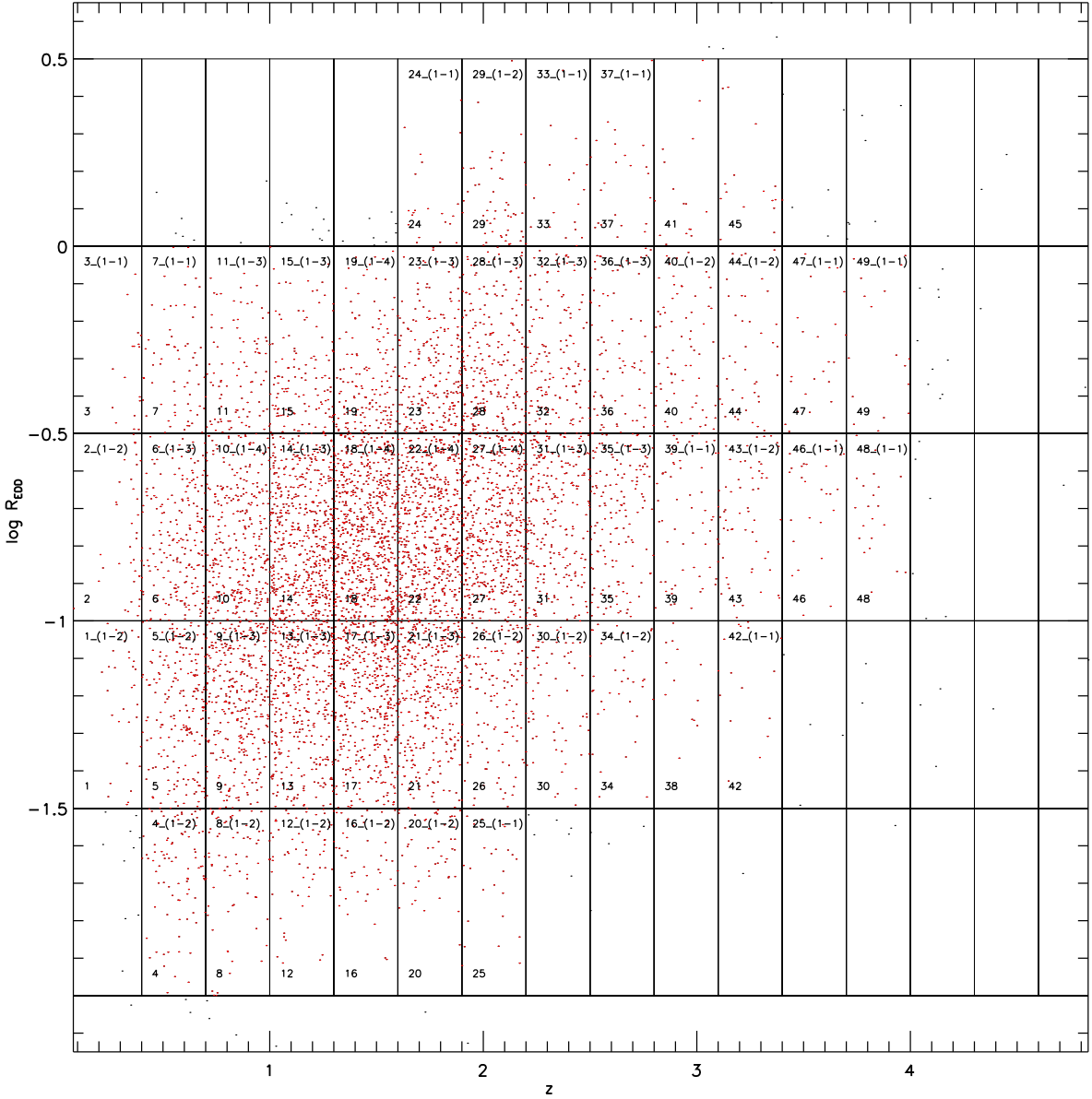}
\caption{$\log{\Redd}$ vs. $z$ for the 7242 radio-quiet quasars.
The bin size of $\Redd$ and $z$ are 0.5 dex and 0.3 respectively. Lines
over-plotted divide quasars into 16$\times$5 subsamples. The 49 grids filled with red
points refer to the qualified subsamples and they are marked with algebraic numbers
in the lower corner of each grid. They are further separated into sub-subsamples with
$\log{\Mbh}$ bin size of 0.5 dex. The algebraic numbers in the upper corner of the grids represent 
the indices of qualified sub-subsamples (with more than 10 quasars in each) for the study of 
the dependence of the variability amplitude on $\lamrf$. 
\label{binlam}}
\end{figure}

To isolate the dependence of the variability amplitude on rest-frame wavelength for
the radio-quiet quasars, we first divide the 7242 radio-quiet quasars into
small bins in a two-dimensional space of redshift and Eddington ratio,
as shown in Fig.~\ref{binlam}. 
Ranging from the minimum value to the maximum value, $z$ is divided into 16 bins. 
$\log{\Redd}$ is divided into 5 bins, as listed in Table~\ref{parambin}. This procedure 
yields 16$\times$5 possible subsamples. Each subsample is further divided in 
6 $\log {\Mbh}$ bins. As seen in Fig.~\ref{binlam}, 49 subsamples containing more than 
10 quasars in each, which are filled with red points, are further subdivided in $\log {\Mbh}$. 
The indices of the 49 qualified subsamples are shown in the lower corner of the corresponding grids. 
104 qualified sub-subsamples with more than 10 quasars in each are finally constructed,
e.g., indices like $1\_$(1-2) means that 2 qualified sub-subsamples are selected from
6 small bins after the division of $\log {\Mbh}$ of the subsample
($0.08<z\le0.4$ and $-1.5<\log \Redd\le-1.0$).
We calculate $r1\_\lamrf$, $r2\_\lamrf$ and fit all the data in each qualified sub-subsample
with the function like Eq.~(\ref{lamequation}).

\begin{figure}[]
\plotone{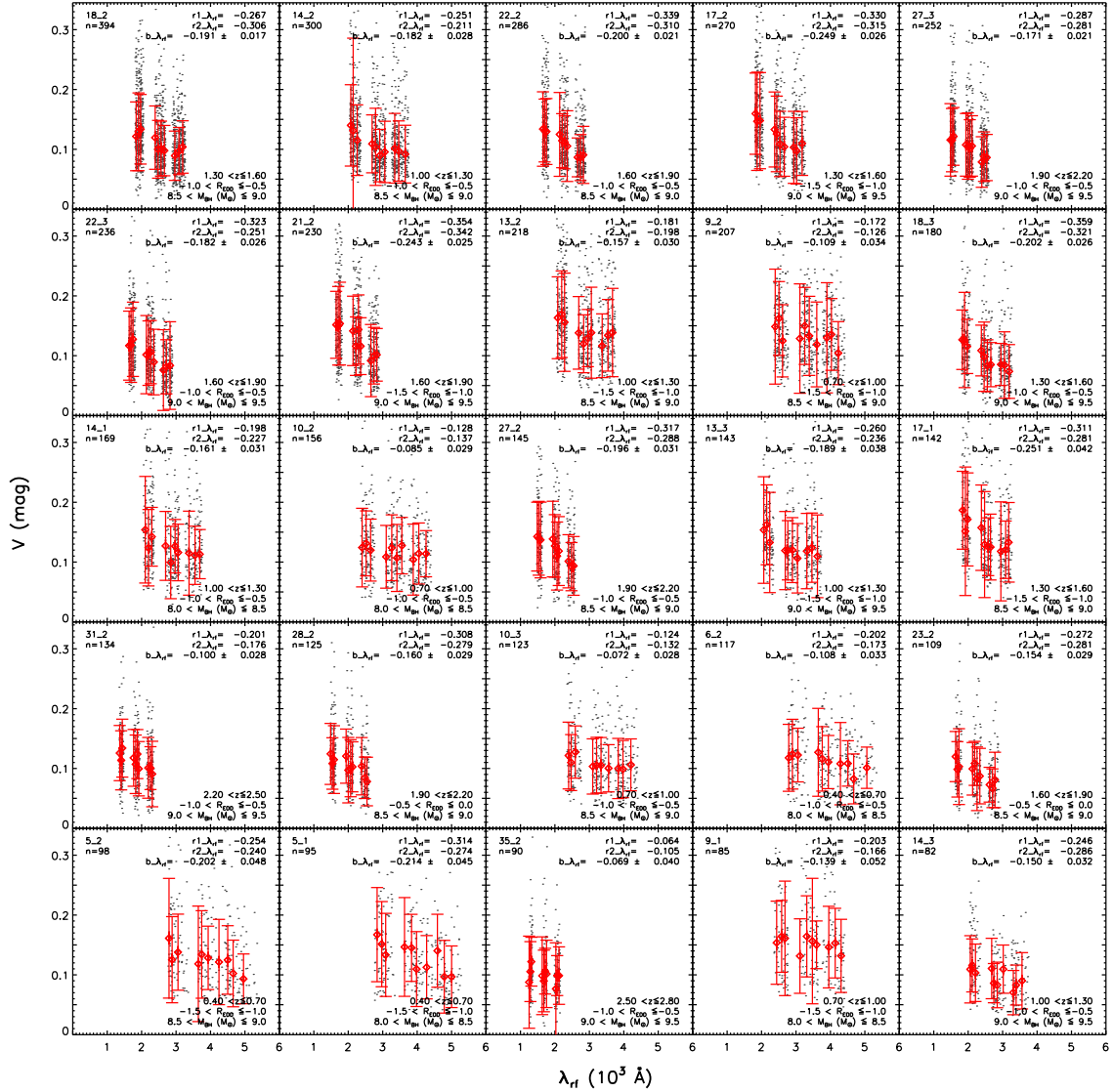}
\caption{Dependence of $V$ on $\lamrf$ in the $g^\prime$, $r^\prime$ and $i^\prime$ band.
To demonstrate the relation, only that for the 25 largest qualified sub-subsamples obtained from
Fig.~\ref{binlam} are shown.
In each panel, the 10 red diamonds refer to the median values of $V$ and $\lamrf$ in the bins
obtained from evenly dividing $\lamrf$ of the sub-subsample.
The legends in the upper left corner of each panel show the index of the sub-subsample
obtained from Fig.~\ref{binlam} and the number of quasars it contains.
$r1\_\lamrf$, $r2\_\lamrf$ and $b\_\lamrf$ are displayed in the upper right corner and
ranges of restricted parameters are shown in the lower right corner.
For the radio-loud quasars, refer to Fig. 21 in the Appendix.
\label{sigma_lamdarq}}
\end{figure}

\begin{figure}[]
\plotone{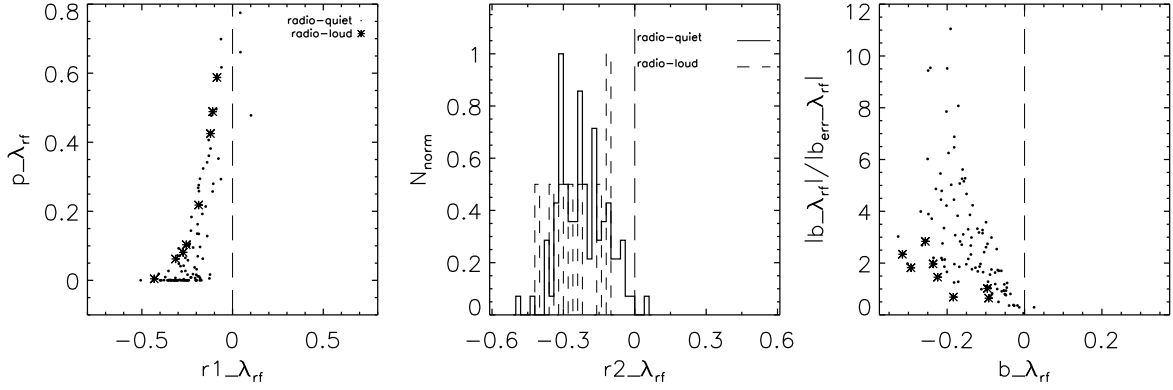}
\caption{Statistical results for the relationships between $V$ and $\log{\lamrf}$:
$p\_{\lamrf}$ vs $r1\_{\lamrf}$(left panel), $r2\_{\lamrf}$ (middle panel) and the linear fit results,
$|b\_{\lamrf}|/|b_{\rm err}\_{\lamrf}|$ vs $b\_{\lamrf}$ (right panel).
Dots and solid lines refer to results of the radio-quiet subsamples,
while asterisk signs and dashed lines refer to results of the radio-loud ones.
\label{zdata_lamda}}
\end{figure}

To demonstrate the rest-frame wavelength $\lamrf$ dependence of the variability amplitude $V$
with significant statistics, in Fig.~\ref{sigma_lamdarq} we show the results of the
25 sub-subsamples containing the most numerous quasars, starting with the largest
sub-subsample.
To make the relation more obvious in each panel, $\lamrf$ of the qualified sub-subsample
is evenly divided into 10 bins. The median variability in each bin is denoted as red diamonds.
Error bars show the standard deviation of variability values for quasars in each rest-frame
wavelength bin. For the radio-loud quasars, we utilize the same mechanism to investigate the
relationship between $V$ and $\lamrf$. 8 qualified sub-subsamples are obtained from binning the
416 radio-loud quasars in $\log{\Redd}$, $z$ and $\log{\Mbh}$. The ranges of these binned parameters 
and the correlations between $V$ and $\lamrf$ are shown in Fig.~\ref{sigma_lamdarl} in the Appendix. 
Note that for radio-loud quasars, figures like Fig.~\ref{sigma_lamdarl}, which describe the dependencies
of variability on the quasar parameters will be shown in the Appendix. For both these
radio-quiet and radio-loud quasars, the anti-correlation between $V$ and $\lamrf$ is evident
---quasars appear to be less variable at larger rest-frame wavelengths. It is further
confirmed by the distributions of calculated $r1\_\lamrf$, $r2\_\lamrf$, $b\_\lamrf$
and $b_{\rm err}\_\lamrf$ values , as shown in Fig.~\ref{zdata_lamda}.
In the left panel, the dots on the left of $r1\_\lamrf=0$ suggest that the majority of the qualified
radio-quiet sub-subsamples exhibit negative $r1\_\lamrf$, with a median value of -0.229.
The significance $p\_\lamrf$ of its value deviation away from 0 is 0.019, which confirms
the moderate anti-correlation at a confidence level over 98\%, while the distribution of the asterisk
signs presents that the median $r1\_\lamrf$ value for all the 8 radio-loud sub-subsamples is
-0.187 with a lower significance level (p$\sim$0.218).
In the middle panel, $r2\_\lamrf$ values for the 100 radio-quiet and all the 8 radio-loud
sub-subsamples are smaller than 0. The median values are -0.233 and -0.224 for the radio-quiet and
the radio-loud data sets respectively, while the distribution of the radio-quiet quasars is broader
than the radio-loud data sets. From the fitting procedure, the median values of $b\_{\lamrf}$ and
$a\_{\lamrf}$ for the radio-quiet sub-subsamples are  -0.137$\pm$0.052 and 0.579$\pm$0.180, while
the median $b\_\lamrf$ and $a\_\lamrf$ value for the radio-loud sub-subsamples are
-0.225$\pm$0.143 and 0.950$\pm$0.494. The relative ratios of $b\_\lamrf$ and their errors
$b_{\rm err}\_\lamrf$ are shown in the right panel, where the ratios are generally larger
for these radio-quiet sub-subsamples, implying a more robust linear relation.

\section{Dependence on redshift\label{redshift}}
\begin{figure}[]
\plotone{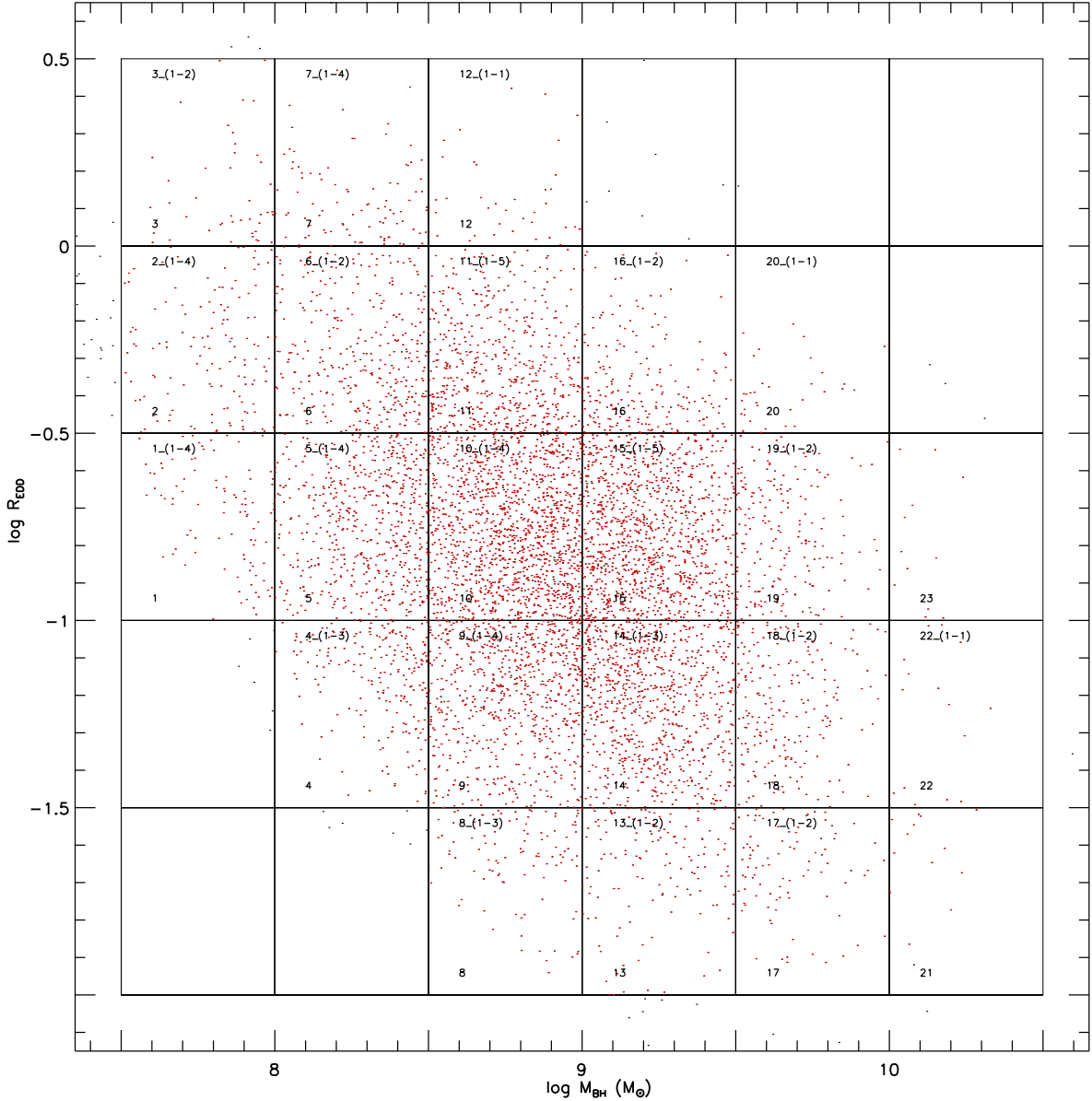}
\caption{$\log{\Redd}$ vs. $\log{\Mbh}$ for the 7242 radio-quiet quasars. Their
bin size are both 0.5 dex. Among 5$\times$6 subsamples divided by the
overdrawn lines, 23 subsamples with more than 10 quasars in each (marked with algebraic numbers
in the lower corner of each grid) are further divided into sub-subsamples with the $\lamrf$ bin size of 
400 \AA (see \textsection~\ref{redshift} for more details). The algebraic numbers in the upper corner of 
the grids refer to the indices of the qualified sub-subsamples for the study of the dependence of $V$ on $z$.
\label{binz}}
\end{figure}

\begin{figure}[]
\plotone{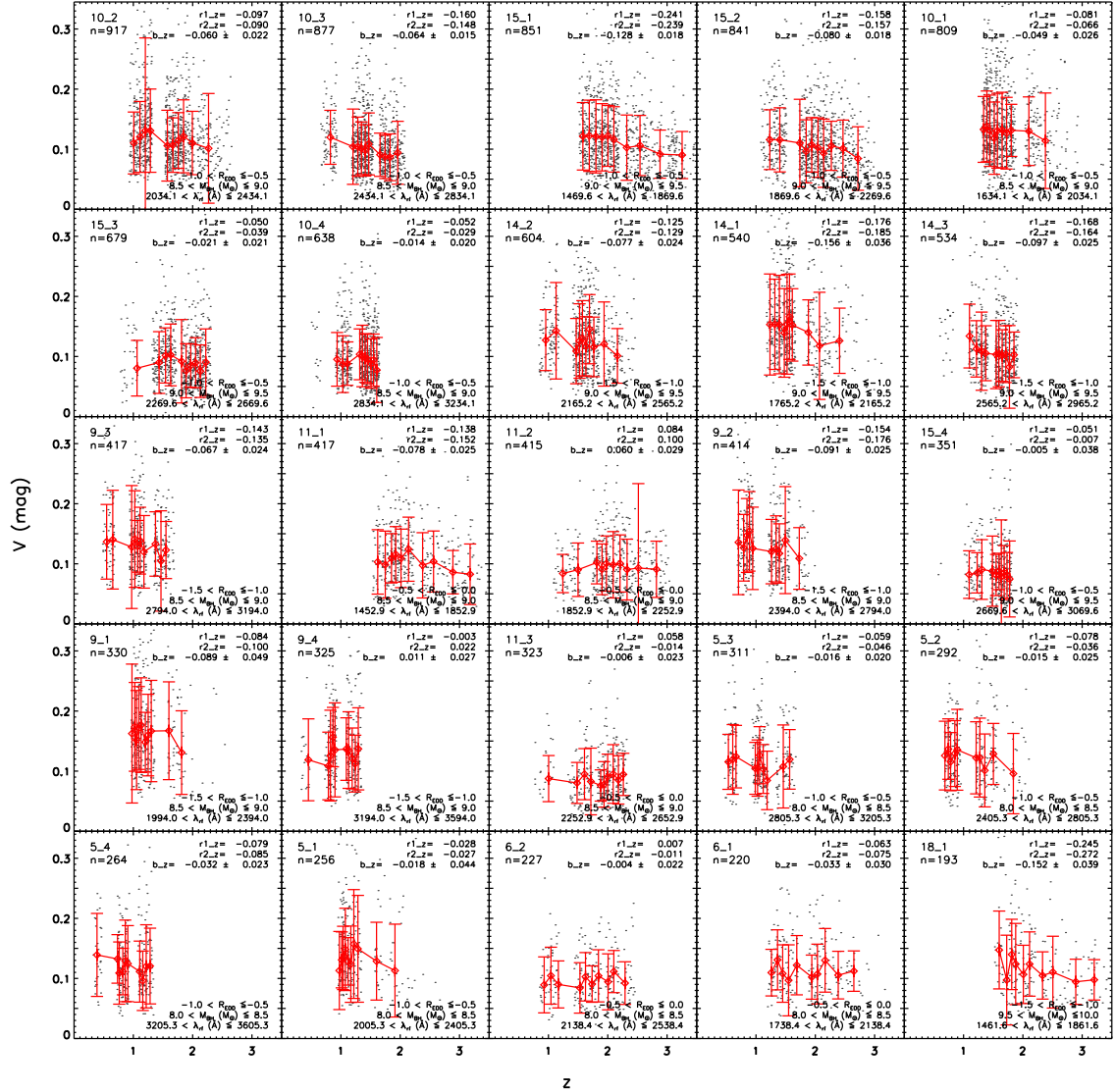}
\caption{Dependence of $V$ on $z$ in the $g^\prime$, $r^\prime$ and $i^\prime$ band for the
25 largest qualified sub-subsamples extracted from Fig.~\ref{binz}.
In each panel, the 10 diamonds refer to the median values of $V$ and
$z$ in bins from evenly dividing $z$.
The texts in each panel have the same meaning with Fig.~\ref{sigma_lamdarq}.
For the radio-loud quasars, refer to Fig. 22 in the Appendix.
\label{sigma_zrq}}
\end{figure}

    We first exclude the influences of the intrinsic quasar parameters on the variability
amplitude by dividing the radio-quiet and radio-loud quasar sample into small bins in
the $\log{\Mbh}$-$\log{\Redd}$ space with their bin sizes of 0.5 dex.
For radio-quiet quasars, 23 subsamples are considered qualified, with
more than 10 quasars in each one. As shown in Fig.~\ref{binz}, 23 grids marked with algebraic numbers 
in the lower side of the grids and filled with red points refer to those qualified subsamples.
For the radio-loud quasars, only 11 subsamples are qualified.
To eliminate the influence of rest-frame wavelength on the relation between the variability
amplitude and redshift, we prefer to divide $\lamrf$ into small bins. However, as the
rest-frame wavelength and redshift are closely related, the binning of $\lamrf$ in one individual
SDSS filter band will lead to a decrease of the redshift range, causing the evaluated dependence of
variability on redshift unreliable \citep{VB04}. Therefore, we treat variability amplitudes in
the 3 bands ($g^\prime$, $r^\prime$ and $i^\prime$) for an individual quasar as variability
amplitudes for 3 individual quasars with the same redshift. Based on the variability amplitude
and $\lamrf$ of the triple-enlarged subsample, we choose an intersecting range of $\lamrf$ in
the 3 bands for quasars in each qualified subsample, which are binned with an equal size of 400 \AA.
In this way, we manage to exclude the influence
of $\lamrf$ on the correlation between $V$ and $z$ by restricting the sub-subsample in small
ranges of rest-frame wavelength, but without decreasing the range of $z$.
We use the same method mentioned in \textsection~\ref{wavelength} to select qualified
sub-subsamples, e.g., indices like 1$\_$(1-4) in Fig.~\ref{binz} means that after the $\lamrf$
division of the subsample ($10^{7.5}$ M$_{\odot} < \Mbh \le 10^{8}$ M$_{\odot}$ and
$-1 < \log{\Redd} \le -0.5$) into several sub-subsamples, 4 of them are qualified with more than 10
quasars in each. The ranges of $\Mbh$ and $\Redd$ for each sub-subsample are shown in Fig.~\ref{binz}
and the range of $\lamrf$ is shown further in the lower right of each panel in Fig.~\ref{sigma_zrq}.

\begin{figure}[]
\plotone{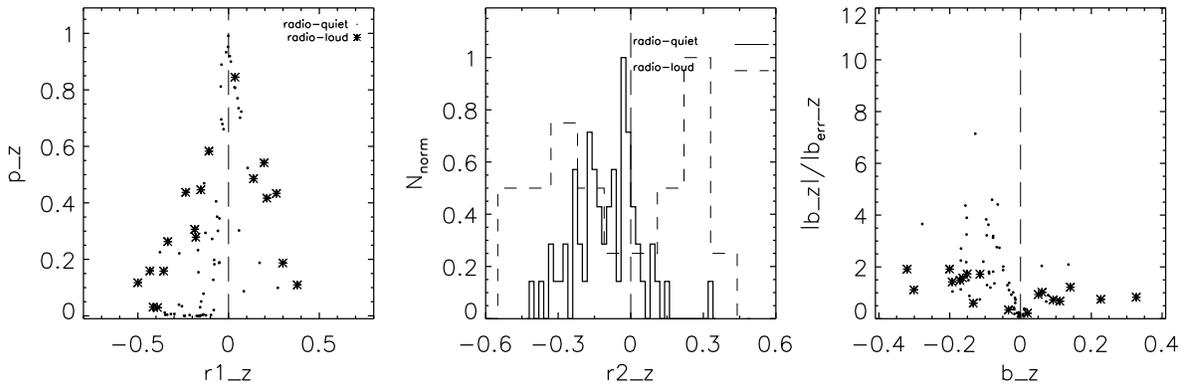}
\caption{Figure similar to Fig.~\ref{zdata_lamda}, but for the statistical results of the 
relationship between $V$ and $\log{z}$.
\label{zdata_z}}
\end{figure}

    The relation between the variability amplitude and redshift for the 25 qualified
radio-quiet sub-subsamples containing the largest numbers of quasars are shown in Fig.~\ref{sigma_zrq},
including the values of $r1\_z$, $r2\_z$, $b\_z$ and the ranges for all the restricted parameters.
We evenly separate the quasars into 10 redshift bins, to make these tendencies more obviously seen.
The median variability amplitude $V$ and redshift $z$ in each bin
are denoted as red diamonds, and error bars show the dispersion of variability values
for these quasars. These results are fairly noisy, and it is difficult to detect any clear
trend with redshift, which is further confirmed in Fig.~\ref{zdata_z}.
Again in a similar way, the 11 qualified radio-loud subsamples are divided into
sub-subsamples with a bin size of rest-frame wavelength of 400 \AA. 18 qualified sub-subsamples
are obtained and details of their redshift dependence of variability are shown in
Fig.~\ref{sigma_zrl} in the Appendix. 

$r1\_z$, $p\_z$, $r2\_z$ and the results from the linear fitting for both the radio-quiet
and radio-loud data sets are shown in Fig.~\ref{zdata_z}. 78.3\% of the radio-quiet sub-subsamples
show negative $r1\_z$ and $r2\_z$ values. However, as 42.6\% of the negative $r1\_z$ values for
the radio-quiet ones are close to zero within a range of 0.1, the median $r1\_z$ value is -0.081
above 81.1\% confidence level, suggesting almost no relation.
Meanwhile, the middle panel indicates that no linear relation between them exists, consistent
with the distribution of $|b\_z|/|b_{\rm err}\_z|$ in the right panel---65.0\% of these data sets show
$|b\_z|/|b_{\rm err}\_z|$ values less than 2.
For the radio-loud data sets, 61.1 percent show negative relations; the median $r1\_z$ value is -0.153 
with the median confidence level ($p\_z \sim$ 0.306). The very weak negative relation has to be concerned
with a larger radio-loud data set due to its low confidence level. Distributions of $r2$ and 
$|b\_z|/|b_{\rm err}\_z|$ show no linear relation between them---55.6 percent of these 
sub-subsamples exhibit an anti-correlation between $V$ and $z$, while the other ones show a positive 
relation.

\section{Dependence on black hole mass, luminosity and Eddington ratio\label{param}}
\begin{figure}[]
\plotone{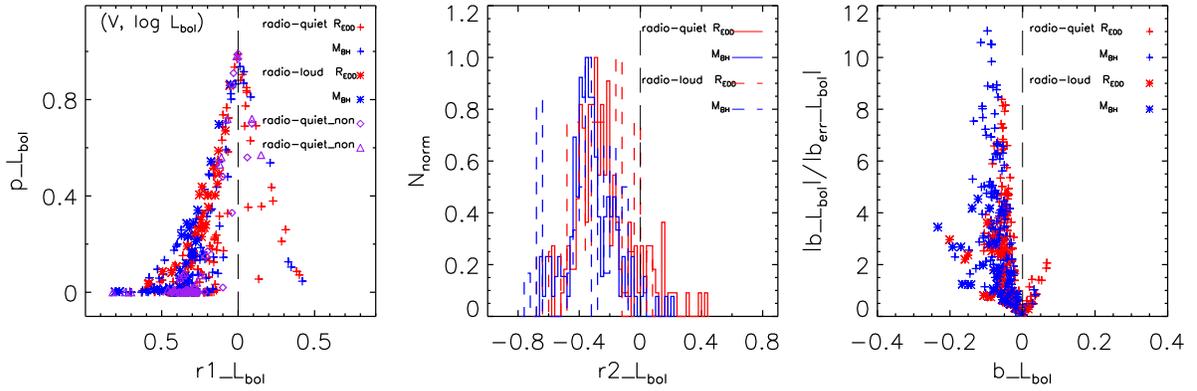}
\caption{Figure similar to Fig.~\ref{zdata_lamda}, but for the statistical results of the relationship
between $V$ and $\log{\Lbol}$. The magenta diamonds and triangles show respectively the results of 
the radio-quiet and radio-loud subsamples. These subsamples are only binned in redshift, thus 
without excluding the relations among the three intrinsic quasar parameters.
The blue signs and lines represent results of the subsamples obtained from the $\log{\Mbh}$-$z$
space and red ones refer to results of the subsamples binned in the $\log{\Redd}$-$z$ space.
\label{zdata_lbol}}
\end{figure}

\begin{figure}[]
\plotone{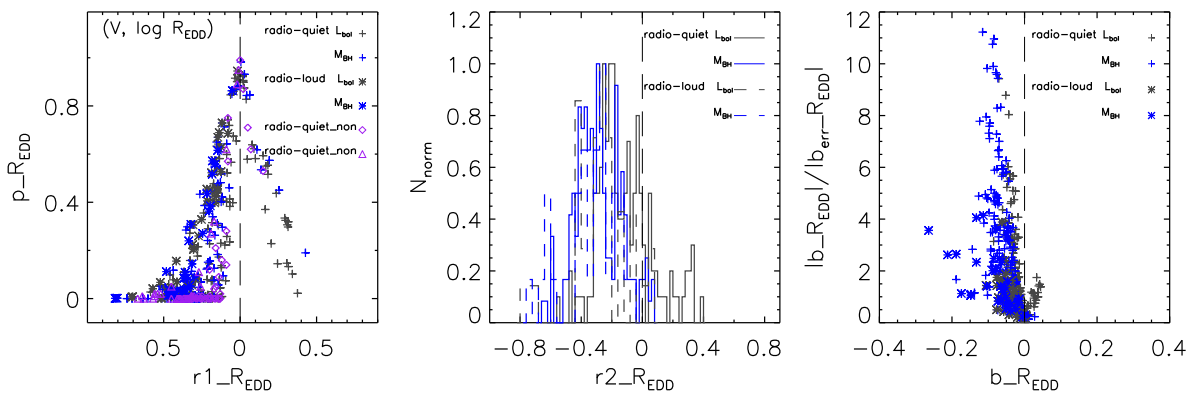}
\caption{Figure similar to Fig.~\ref{zdata_lbol}, but for the statistical results of the relationships
between $V$ and $\log{\Redd}$.
The grey signs and lines represent results of the subsamples obtained from the $\log{\Lbol}$-$z$ space
and blue ones represent results of the subsamples obtained from the $\log{\Mbh}$-$z$ space.
\label{zdata_edd}}
\end{figure}

\begin{figure}[]
\plotone{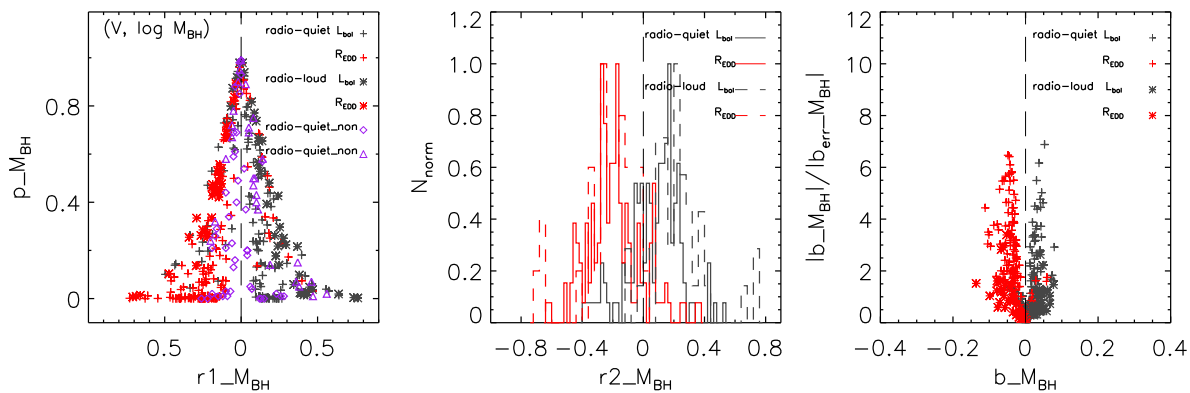}
\caption{Figure similar to Fig.~\ref{zdata_lbol}, but for the statistical results of the relationship
between $V$ and $\log{\Mbh}$.
The red signs and lines represent the results from the subsamples obtained from the $\log{\Redd}$-$z$ space
and grey ones represent the results from the subsamples obtained from the $\log{\Lbol}$-$z$ space.
\label{zdata_mbh}}
\end{figure}

Both the radio-quiet and radio-loud sample are first divided into 16 redshift bins as mentioned
in \textsection~\ref{wavelength}.
For the 14 qualified subsamples of radio-quiet quasars with redshift ranging from 0.4 to 4.3 and
9 subsamples of the radio-loud quasars with redshift spanning from 0.4 to 3.1, we calculate
Spearman's Rank Correlation Coefficients between the variability amplitude in the 3 SDSS bands
($g^\prime$, $r^\prime$ and $i^\prime$) and the three intrinsic quasar parameters
($\Mbh$, $\Lbol$ and $\Redd$), without accounting for the relationships among the intrinsic quasar
parameters themselves. Results for the radio-quiet and radio-loud quasar subsamples are shown as magenta
diamonds and triangles in the left panel of Fig.~\ref{zdata_lbol}, Fig.~\ref{zdata_edd} and
Fig.~\ref{zdata_mbh}, respectively.

For both the radio-loud and radio-quiet subsamples, there is a weak anti-correlation
between variability and $\Redd$ at a high confidence level. To make a distinction
between the Spearman's Rank Correlation Coefficients here and $r1$ calculated after
considering relationships among the intrinsic quasar parameters, we denote results here as the
format like $r1^n\_X$. $r1^n\_\Redd$ of 12 radio-quiet and all the radio-loud subsamples are
less than 0 and the median values for these radio-quiet and radio-loud subsamples are
approximately -0.170 and -0.390 respectively. Most of the $p^n\_\Redd$ values are around 0. A weak
anti-correlation between the variability amplitude and $\Lbol$ shows up as more than 90\% subsamples
exhibit negative $r1^n\_\Lbol$ values at a confidence level over 99\%. The median $r1^n\_\Lbol$
values for the radio-quiet and the radio-loud subsamples are -0.290 and -0.350 respectively.
However, no conclusive results are found for the relation with $\Mbh$.
This is consistent with the anti-correlation between variability and $\Lbol$ found
in previous studies on smaller samples, where the relations among the intrinsic quasar
parameters were not taken into consideration either \citep{Cristiani96, Di96, Paltani97}.
The correlation between variability and $\Mbh$ for the $0.08<z\le0.7$ radio-quiet sample
($r1^n\_\Mbh$ $\sim$ 0.26, $p^n\_\Mbh$ $\sim$ 0.01 for the quasars within $0.08<z\le0.4$ and
$r1^n\_\Mbh$ $\sim$ 0.13, $p^n\_\Mbh$ $\sim$ 0.01 for the quasars within $0.4<z\le0.7$), corresponds
well to the significant dependence of variability on $\Mbh$ for the $0<z<0.75$ sample
reported by \cite{Wold07}. We discuss this further in \textsection~\ref{discussion}.

     However, the relations among the intrinsic quasar parameters have to be taken into
account to investigate the influences of different parameters on the variability
amplitude. As more luminous quasars tend to posses more massive black holes,
the relationship between variability and black hole mass will also express itself in
the correlation between variability and luminosity or the relation with
Eddington ratio. As shown in this section, after considering the relationships among the 
intrinsic quasar parameters, trends are still clear in some cases but become ambiguous or fade 
out in other cases.
The radio-quiet quasars and radio-loud quasars are subdivided into small data sets
by binning redshift and one of the three quasar parameters to exclude their influences on
the relationships between the variability amplitude and the other two quasar parameters.
The separate consideration of variability amplitudes in the 3 filter
bands manage to restrict the rest-frame wavelength of each data set to small ranges.

\begin{figure}[]
\plotone{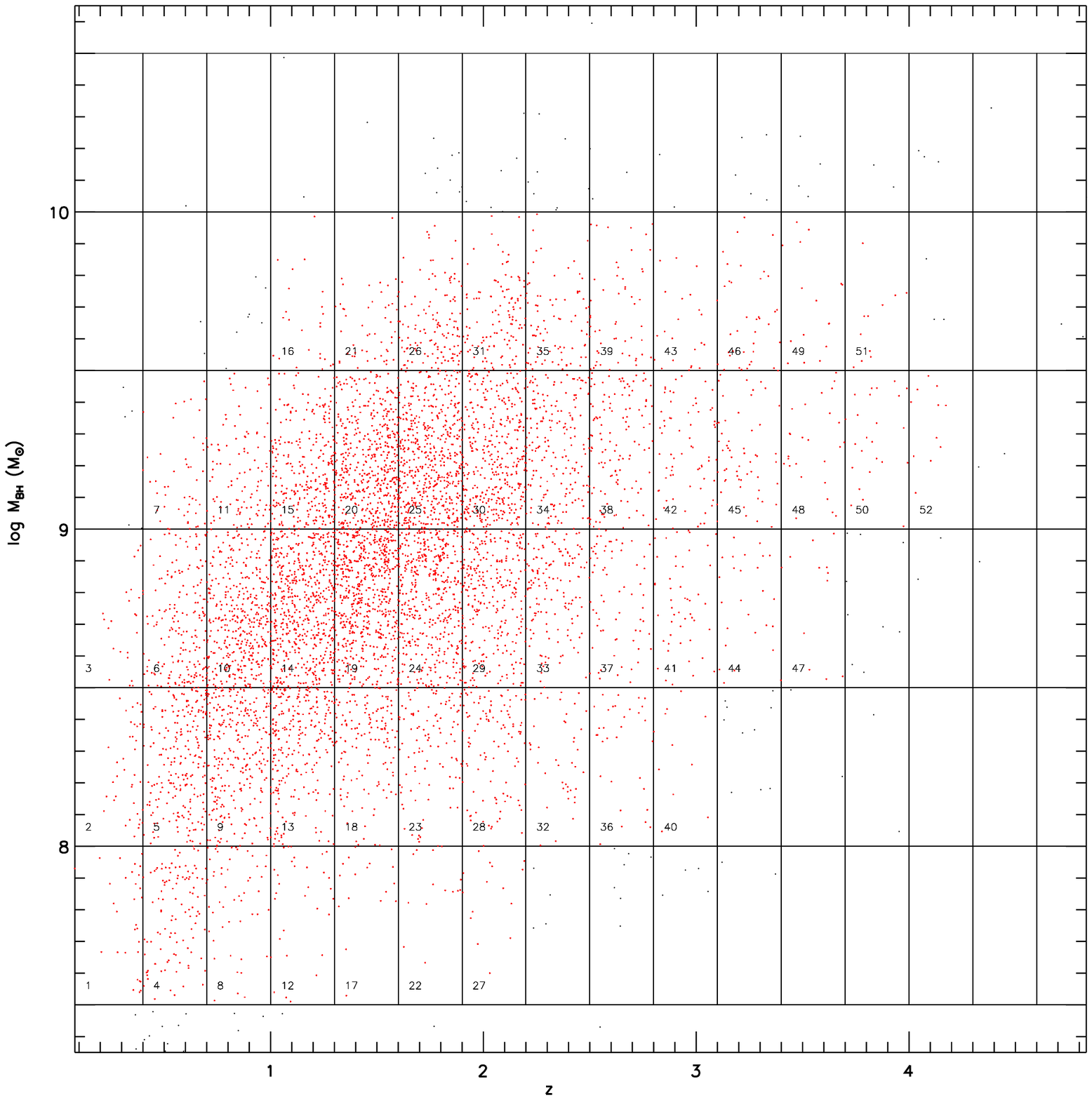}
\caption{$\log{\Mbh}$ vs. $z$ for the 7242 radio-quiet quasars. Their bin sizes
are 0.5 dex and 0.3 respectively. Among 6$\times$16 subsamples divided by
overdrawn lines, 52 subsamples with more than 10 quasars in each are qualified,
represented by the grids filled with red points and marked with algebraic numbers.
\label{binmbh}}
\end{figure}

\begin{figure}[]
\plotone{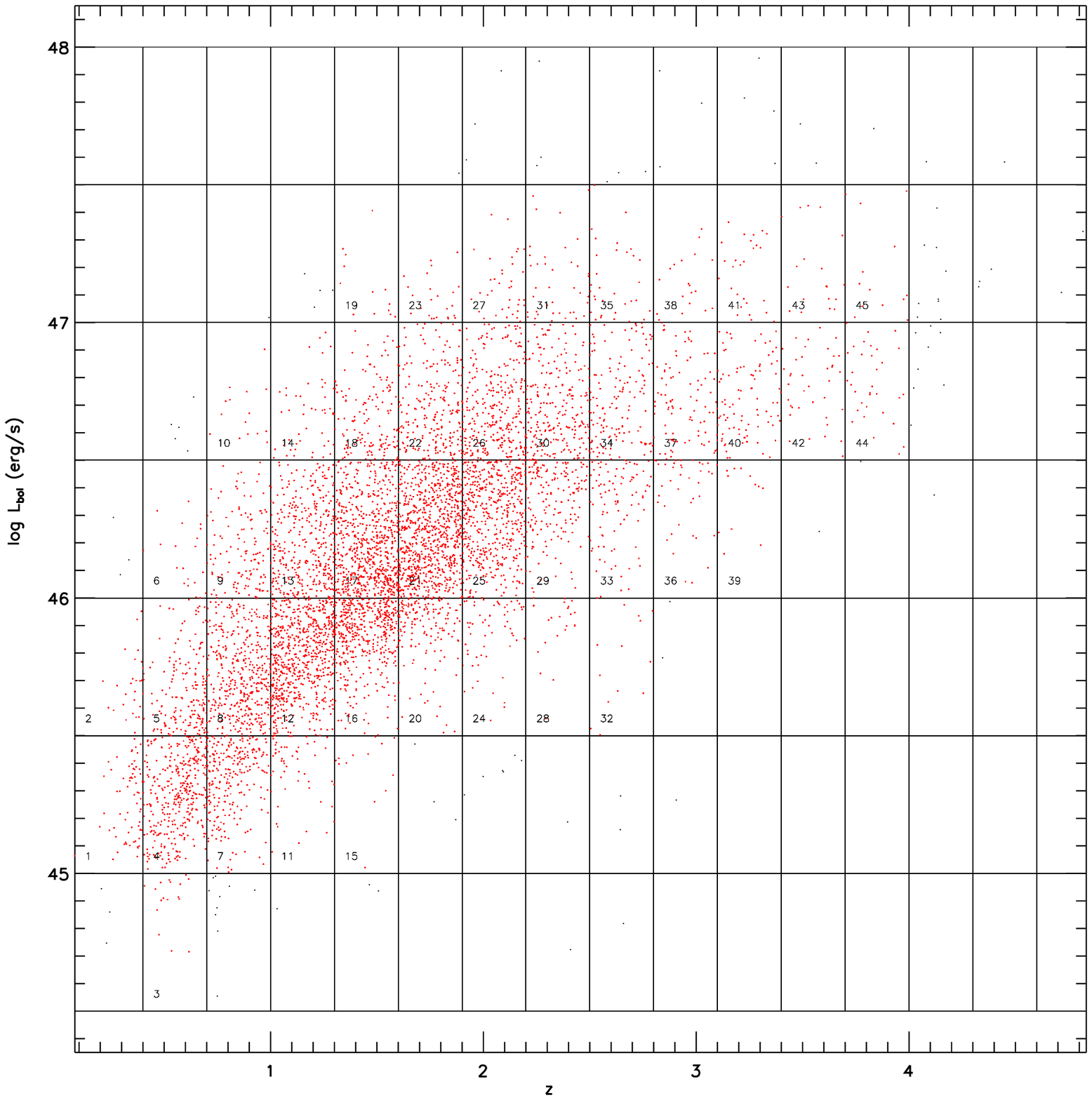}
\caption{$\log{\Lbol}$ vs. $z$ for the 7242 radio-quiet quasars. Their bin sizes
are 0.5 dex and 0.3 respectively. Among 7$\times$16 subsamples divided by
overdrawn lines, 45 subsamples with more than 10 quasars in each are qualified,
represented by the grids filled with red points and marked with algebraic numbers.
\label{binlbol}}
\end{figure}

    Fig.~\ref{binlam} shows the details of binning in the $\log{\Redd}$-$z$ space for
the radio-quiet quasars and yields 49 qualified subsamples, as described in
\textsection~\ref{wavelength}.
Fig.~\ref{binmbh} and Fig.~\ref{binlbol} show the details of binning in the
$\log{\Mbh}$-$z$ space and $\log{\Lbol}$-$z$ space, respectively. In Fig.~\ref{binmbh}
, among 16$\times$6 grids, 52 grids refer to the qualified subsamples.
In Fig.~\ref{binlbol}, 16$\times$7 grids are obtained and 45 grids are qualified.
For the radio-loud quasars, 11, 11 and 13 qualified subsamples are extracted in the $\log{\Redd}$-$z$,
$\log{\Mbh}$-$z$ and $\log{\Lbol}$-$z$ spaces, respectively.
In each qualified subsample, we further calculate $r1$, $r2$, $b$ and $a$ for the
correlations between $V$ and the other two unrestricted physical parameters.
Their distributions are shown statistically in Fig.~\ref{zdata_lbol}, ~\ref{zdata_edd} and
~\ref{zdata_mbh}. For the subsamples obtained from binning in the $\log{\Redd}$-$z$ space,
the corresponding signs (plus signs for results from the radio-quiet data sets
and asterisk signs for results from the radio-loud data sets) are colored as red, while for the subsamples
from binning in the $\log{\Mbh}$-$z$ and $\log{\Lbol}$-$z$ space, signs are colored as blue and grey
respectively. Thus in the panels of Fig.~\ref{zdata_lbol}, Fig.~\ref{zdata_edd} and
Fig.~\ref{zdata_mbh}, which contain plus and asterisk signs,
the number of red, blue and grey plus signs are 49$\times$3 (the $g^\prime$, $r^\prime$
and $i^\prime$ filter band), 52$\times$3 and 45$\times$3, while the number of red, blue and
grey asterisk signs are 11$\times$3, 11$\times$3 and 13$\times$3 respectively.

\begin{figure}[]
\plotone{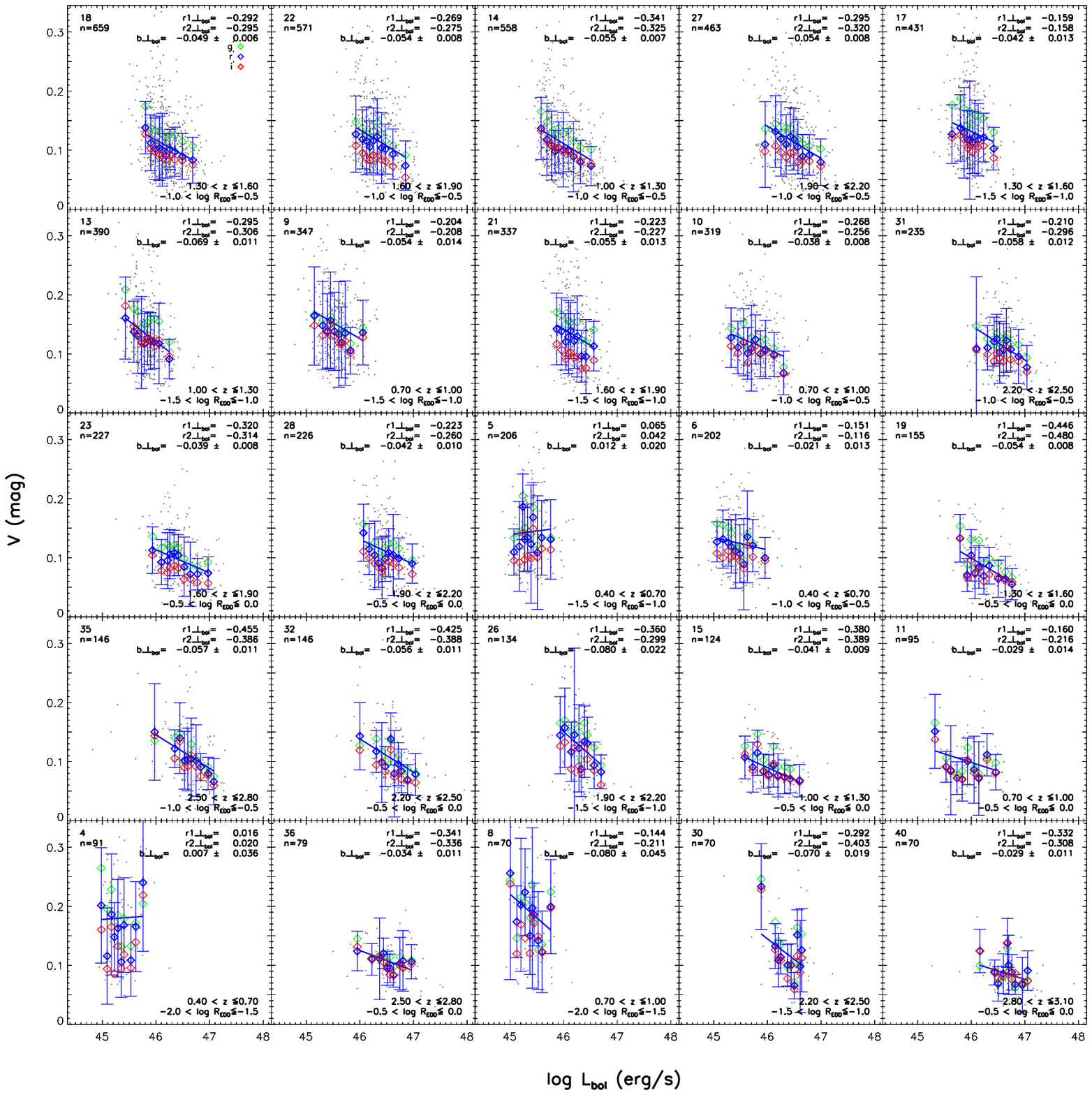}
\caption{Dependence of $V$ in the $g^\prime$, $r^\prime$ and $i^\prime$ filter band on $\Lbol$ for the 25 
largest qualified subsamples obtained from the $\log{\Redd}$-$z$ space as shown in Fig.~\ref{binlam}. 
In each panel, only $V$ in the $r^\prime$ band for all the quasars are represented by grey points. 
The 10$\times$3 colored diamonds (green, blue and red) show the median 
$V$ values in the 3 bands ($g^\prime$, $r^\prime$ and $i^\prime$)
in the bins obtained from evenly dividing $\log{\Lbol}$.
The blue lines represent the linear fits for all the data in the $r^\prime$ band.
The legends in the upper right corner of each panel list the $r1\_\Lbol$, $r2\_\Lbol$
and $b\_\Lbol$ for the study of the variability amplitudes in the $r^\prime$ band, while
the texts in the lower right corner of each panel give the ranges of restricted
parameters. For the radio-loud quasars, refer to Fig. 23 in the Appendix.
\label{sigma_eddratio_lbol}}
\end{figure}

\begin{figure}[]
\plotone{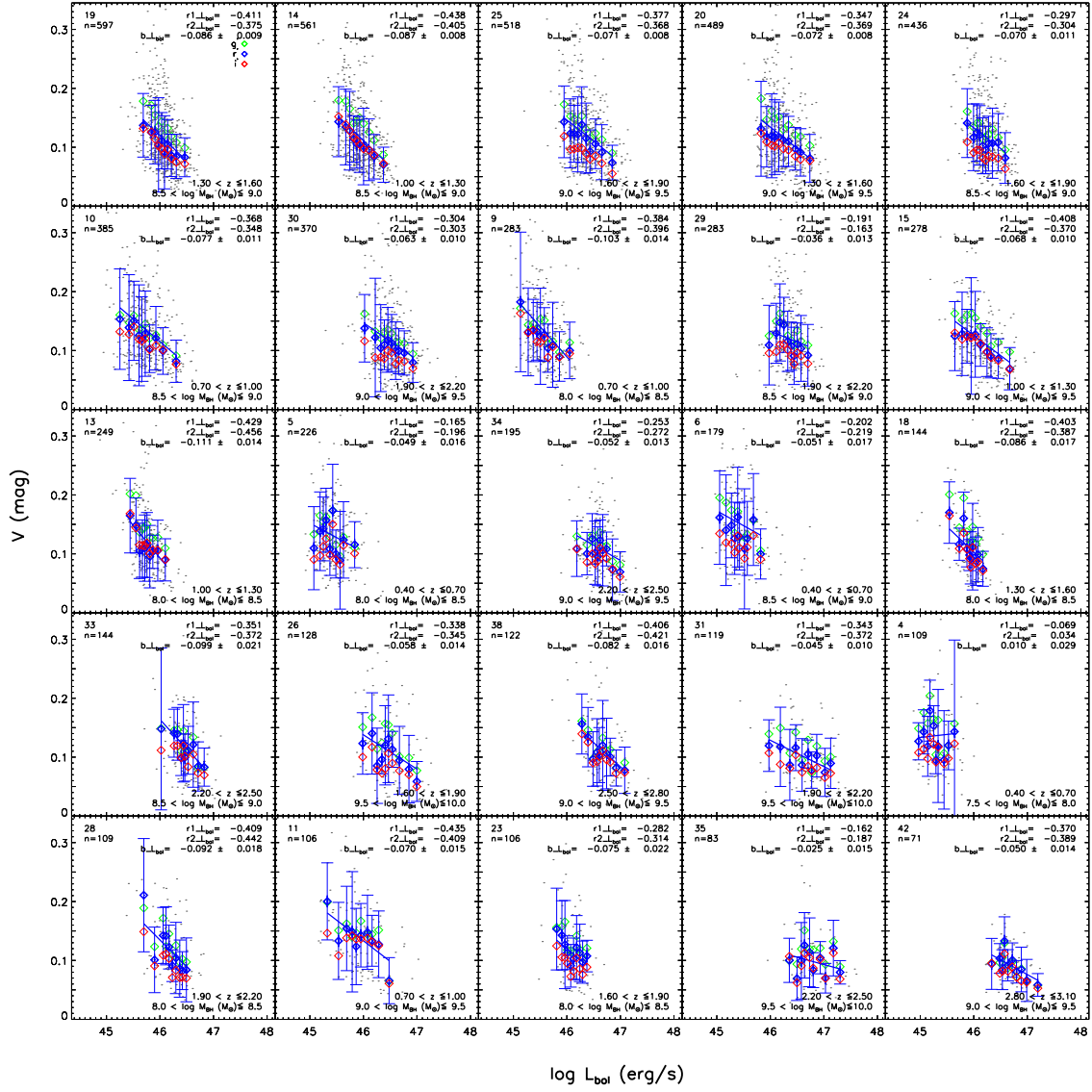}
\caption{Figure similar to Fig.~\ref{sigma_eddratio_lbol}, but for the 25 largest qualified
subsamples obtained from the $\log{\Mbh}$-$z$ space as displayed in Fig.~\ref{binmbh}. Refer to
Fig. 24 in the Appendix with respect to the radio-loud quasars.
\label{sigma_mbh_lbol}}
\end{figure}

To demonstrate the dependence of the variability amplitude on $\Lbol$ in the qualified subsamples
obtained from the $\log{\Redd}$-$z$ space and $\log{\Mbh}$-$z$ space, we further show the detailed
relations for the 25 largest radio-quiet quasar data sets from the former space in
Fig.~\ref{sigma_eddratio_lbol} and that from the latter space in Fig.~\ref{sigma_mbh_lbol}.
To make these underlying tendencies more intuitively clear, in each subsample the unrestricted
parameter $\Lbol$ on which we are studying the dependence of the variability amplitude, is further
subdivided into 10 bins with an equal number of quasars in each one. The median variability amplitudes
and median parameter values are shown as colored diamonds. The error bars represent the 68 percent
dispersion of variability amplitudes for the quasars in each bin. Only in the $r^\prime$ band the
variability amplitudes for all the quasars are further shown as grey points. The fitting of the linear
relation between $V$ and $\log{\Lbol}$ for all the data in each band are carried out, but only
the resulting slope in the $r^\prime$ band is shown in the upper right of each panel.
$r1\_\Lbol$, $p\_\Lbol$ and $r2\_\Lbol$ for the quasars in each subsample in the $r^\prime$ band are also
listed. 94.0 percent of the two figures presents an evident anti-correlation
between $V$ and $\log{\Lbol}$, consistent with the statistical study shown in Fig.~\ref{zdata_lbol}.

As shown in Fig.~\ref{zdata_lbol}, almost all subsamples exhibit an anti-correlation between
the variability amplitude and $\Lbol$ after considering the relations among the intrinsic quasar parameters.
The median value of $r1\_\Lbol$ for the radio-quiet subsamples within small ranges of $z$ and $\log{\Redd}$ is
-0.269 over 98\% confidence level, while the median value of $r1\_\Lbol$ for the radio-quiet subsamples
within small ranges of $z$ and $\log{\Mbh}$ is -0.343 above 99\% confidence level.
The distribution of $r2\_\Lbol$ with a median value of -0.260 and -0.335 for the subsamples
in the $\log{\Redd}$-$z$ space and $\log{\Mbh}$-$z$ space, respectively, as displayed in the middle panel, indicates a
weak linear relation between $V$ and $\log{\Lbol}$. The distribution of the fitting slope $b\_\Lbol$
and its error is shown in the right panel.
Based on the averaged value, the anti-correlation between the variability amplitude and $\Lbol$ can
be described with a format similar to Eq.~(\ref{lamequation}). The median
values of $b\_\Lbol$ and $a\_\Lbol$ for the radio-quiet subsamples in the $\log{\Redd}$-$z$ space are
-0.044$\pm$0.020 and 2.128$\pm$0.913, and for the radio-quiet subsamples in the $\log{\Mbh}$-$z$ space are
-0.063$\pm$0.020 and 3.032$\pm$0.926.
While for the radio-loud subsamples in the $\log{\Redd}$-$z$ space, the median $r1\_\Lbol$, $r2\_\Lbol$,
$b\_\Lbol$ and $a\_\Lbol$ are -0.266 with a lower confidence level ($p\_\Lbol\sim$0.297), -0.269,
-0.058$\pm$0.054 and 2.794$\pm$2.456;
and in the $\log{\Mbh}$-$z$ space, the median $r1\_\Lbol$, $r2\_\Lbol$, $b\_\Lbol$ and $a\_\Lbol$ are -0.384
with a significance of 0.164, -0.321, -0.071$\pm$0.055 and 3.373$\pm$2.536.
As displayed in the right panel, the error in $b\_\Lbol$ is comparable to $b\_\Lbol$ for the
radio-loud quasars, indicating that the dispersion is large and the linear relation is very weak.
According to the simplest Poissonian model, in which starburst events are independent and discrete, 
the luminosity variability is expected to vary with luminosity as 
$\log{V}$ $\propto$ 0.5 $\log{\Lbol}$ \citep{Cid00}.
Despite the possibility that a wide range of $b$ values could be obtained by considering more 
detailed Poissonian models, where supernovae and their remnants were considered, $b$ is often 
close to -0.5\citep{VB04, Paltani97, Aretxaga97}. 
We fit our data with a two-parameter function $\log{V} = b \log{\Lbol} + a$ and the
median $b$ is -0.223$\pm$0.075. Thus, the results obtained from our sample does not support
the Poissonian model.

\begin{figure}[]
\plotone{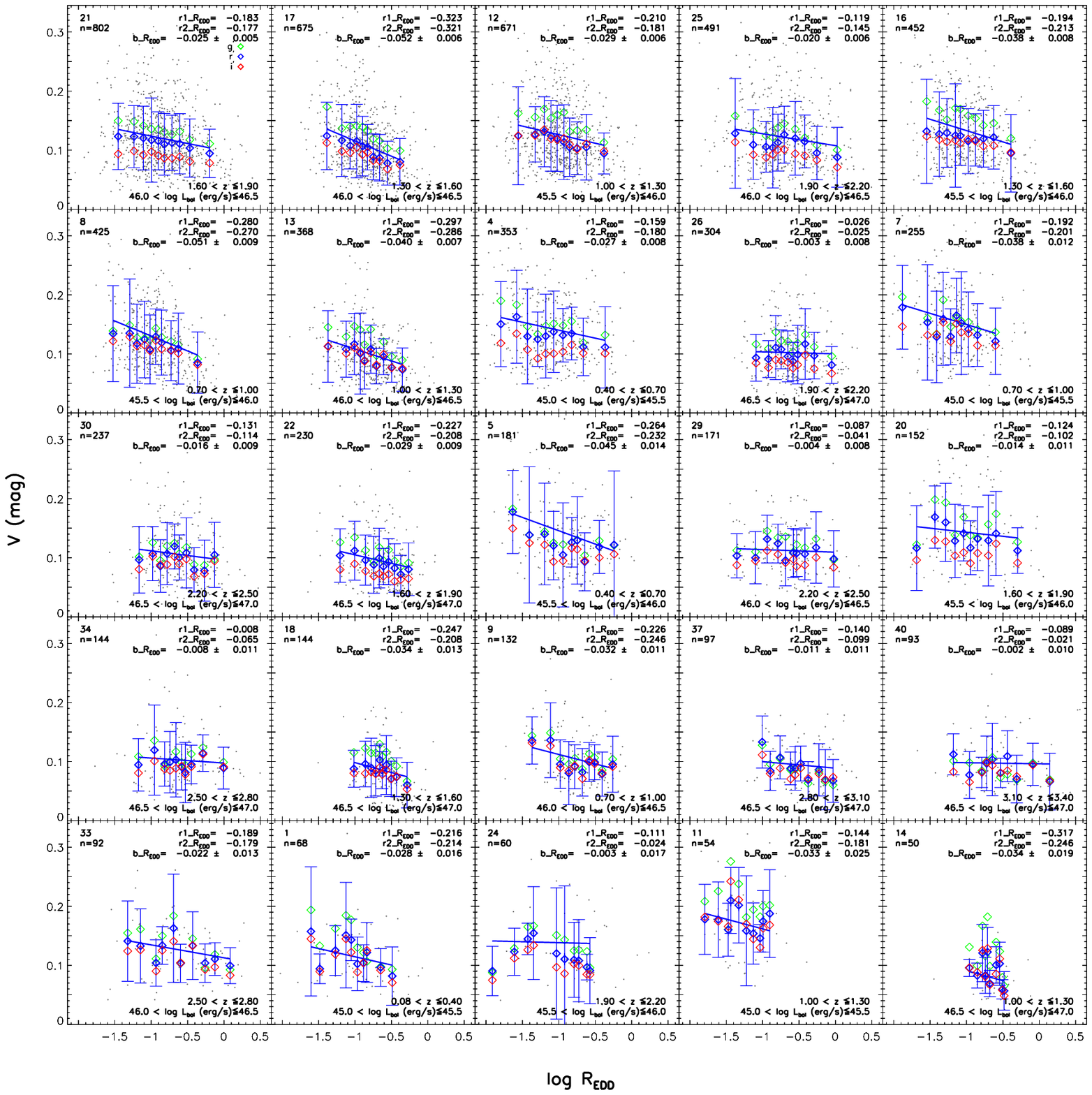}
\caption{Dependence of $V$ in the $g^\prime$, $r^\prime$ and $i^\prime$ filter band on $\Redd$ for the
25 largest qualified subsamples obtained from the $\log{\Lbol}$-$z$ space as shown in Fig.~\ref{binlbol}.
In each panel, only $V$ in the $r^\prime$ band for all the quasars are represented by grey points. 
The 10$\times$3 colored diamonds (green, blue and red) show the median $V$ values in the 3 bands ($g^\prime$, 
$r^\prime$ and $i^\prime$) in the bins obtained from evenly dividing $\log{\Redd}$.
The blue lines represent the linear fits for all the data in the $r^\prime$ band.
The legends in the upper right corner of each panel list the $r1\_\Redd$, $r2\_\Redd$ and $b\_\Redd$ for
the study of the variability amplitudes in the $r^\prime$ band, while the texts in the lower right
corner of each panel show the ranges of the restricted parameters. Refer to Fig. 25
in the Appendix for the radio-loud quasars.
\label{sigma_lbol_eddratio}}
\end{figure}

\begin{figure}[]
\plotone{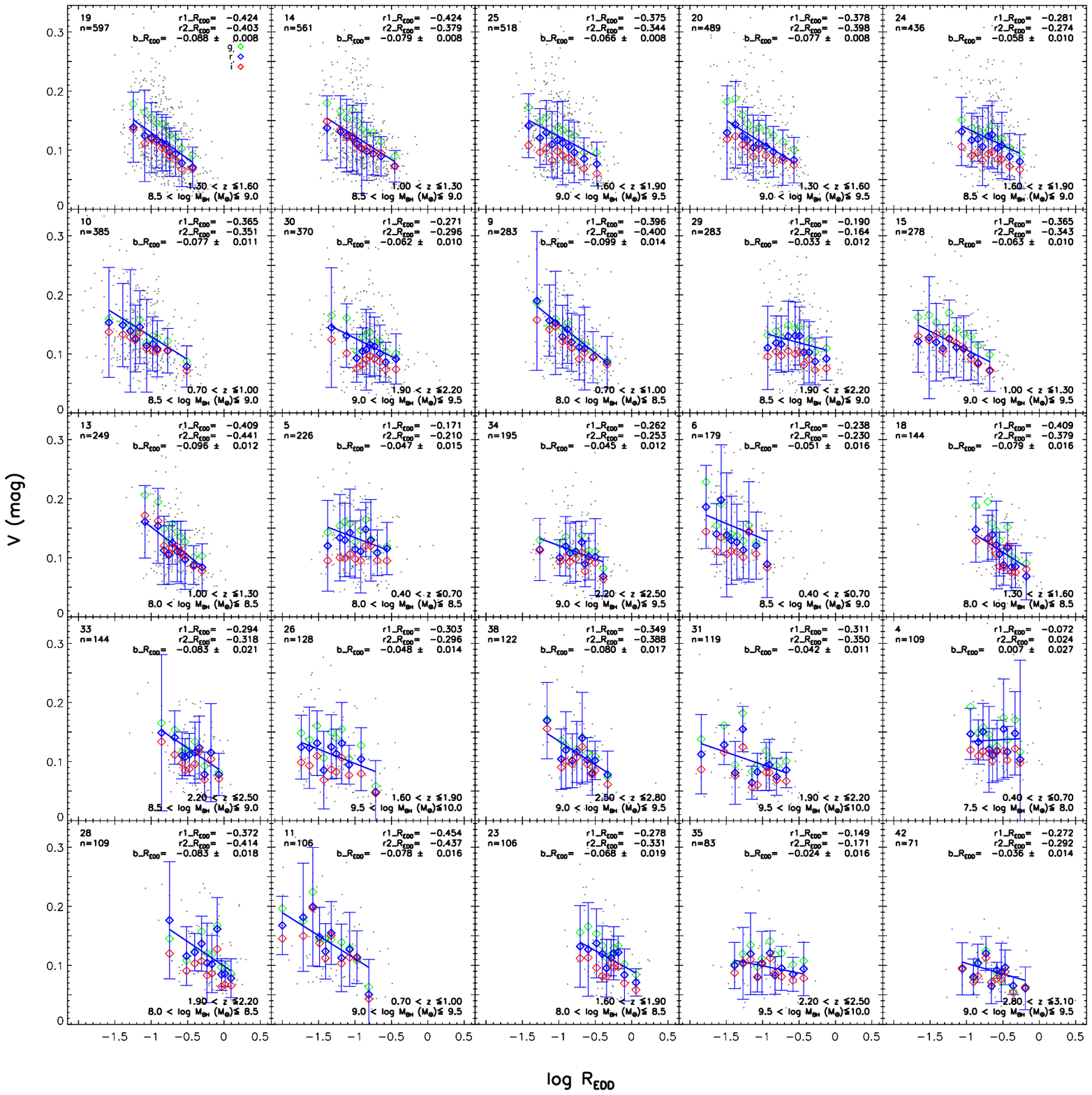}
\caption{Figure similar to Fig.~\ref{sigma_lbol_eddratio}, but for the 25 largest qualified subsamples
obtained from the $\log{\Mbh}$-$z$ space as shown in Fig.~\ref{binmbh}.
Refer to Fig. 26 in the Appendix for the radio-loud quasars.
\label{sigma_mbh_eddratio}}
\end{figure}

The dependence of the variability amplitude on $\Redd$ is analyzed based on the 45 subsamples in the 
$\log{\Lbol}$-$z$ space and 52 qualified subsamples in the $\log{\Mbh}$-$z$ space. To specify
the relation, the details for the 25 largest subsamples in each space are further shown in
Fig.~\ref{sigma_lbol_eddratio} and Fig.~\ref{sigma_mbh_eddratio} respectively. Going from the upper
to the lower panels within various parameter ranges in Fig.~\ref{sigma_mbh_eddratio}, the anti-correlation
between $V$ and $\Redd$ is apparent. While in only 4 panels of Fig.~\ref{sigma_lbol_eddratio},
we find no discernible relation or an insignificant correlation, such as the third panel in the lowest line.

Following the analysis of the relationship between variability and luminosity, statistical results of 
$r1\_\Redd$, $p\_\Redd$, $r2\_\Redd$, $b$ and $|b\_\Redd|/|b_{\rm err}\_\Redd|$ are shown in
Fig.~\ref{zdata_edd}.
For the radio-quiet quasars in the $\log{\Lbol}$-$z$ space and $\log{\Mbh}$-$z$ space, more than 80.0 percent
and 94.2 percent of $r1\_\Redd$ are negative respectively, indicating an anti-correlation between
$V$ and $\Redd$. All the qualified radio-loud subsamples from the two spaces exhibit negative $r1\_\Redd$ values.
Respectively in the two spaces, the median $r1\_\Redd$ value is -0.167 ($p\_\Redd \sim$ 0.137) and
-0.303 ($p\_\Redd \sim$ 0.013) for the radio-quiet quasars; and -0.304 and -0.369 for the radio-loud quasars
over a lower confidence level ($p\_\Redd \sim$ 0.300 and 0.188).
As shown in the last 2 panels, 79.3\% of the qualified radio-quiet subsamples
obtained from the $\log{\Lbol}$-$z$ space exhibit negative $r2\_\Redd$ and $b\_\Redd$ values,
confirming the weak anti-correlation between $V$ and $\log{\Redd}$.
The linear relation for the qualified radio-quiet subsamples binned from the
$\log{\Mbh}$-$z$ space is more evident with a median $r2\_\Redd$ of -0.299. 
More than 92.3 percent of radio-loud subsamples from the $\log{\Lbol}$-$z$ space and
all the qualified ones from the $\log{\Mbh}$-$z$ space display negative linear relations between $V$ and 
$\log{\Redd}$, with median $r2\_\Redd$ values of -0.274 and -0.291 in the $\log{\Lbol}$-$z$ and 
$\log{\Mbh}$-$z$ space respectively. The linear relation is also described with a simple function like 
Eq.~(\ref{lamequation}). $|b\_\Redd|/|b_{\rm err}\_\Redd|$ values are generally larger for the radio-quiet 
subsamples, indicating a stronger linear relation.
The median values of $b\_\Redd$ and $a\_\Redd$ for the radio-quiet subsamples in the $\log{\Lbol}$-$z$ space are
-0.020$\pm$0.014 and 0.091$\pm$0.011, while for the radio-loud quasars the results become
-0.045$\pm$ 0.032 and 0.079$\pm$0.035 respectively. 
For the radio-quiet subsamples in the $\log{\Mbh}$-$z$ space the median values of $b\_\Redd$ and $a\_\Redd$ are
-0.052$\pm$0.019 and 0.075$\pm$0.013; for the radio-loud subsamples they are -0.069$\pm$0.047
and 0.056$\pm$0.055, respectively.

\begin{figure}[]
\plotone{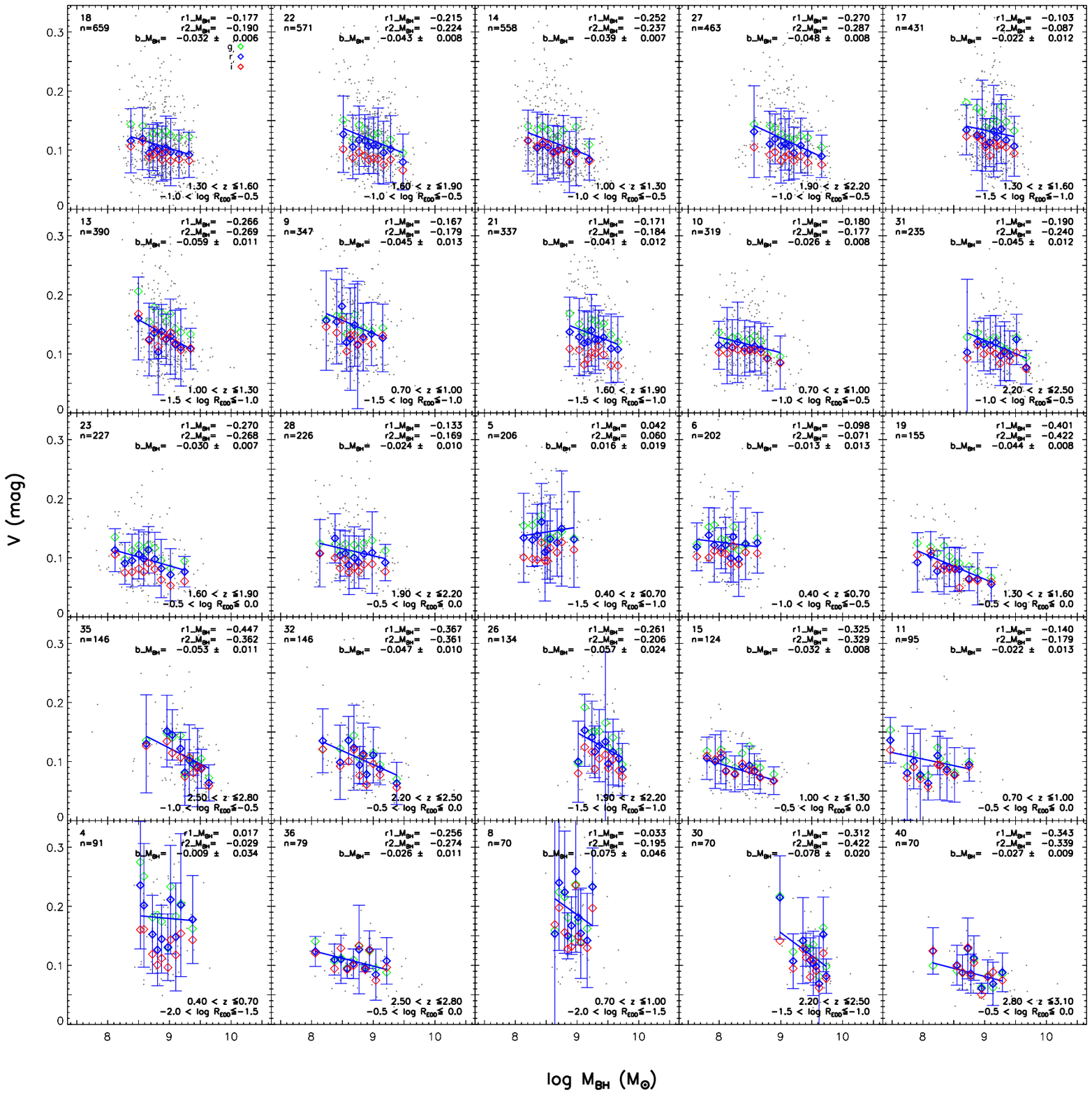}
\caption{Dependence of $V$ in the $g^\prime$, $r^\prime$ and $i^\prime$ filter band on $\Mbh$ for the
25 largest qualified subsamples obtained from the $\log{\Redd}$-$z$ space as shown in Fig.~\ref{binlam}.
In each panel, only $V$ in the $r^\prime$ band for all the quasars are represented by grey points.
The 10$\times$3 colored diamonds (green, blue and red) show the median $V$ values in the 3 bands ($g^\prime$, 
$r^\prime$ and $i^\prime$) in the bins obtained from evenly dividing $\log{\Mbh}$.
The blue lines represent the linear fits for all the data in the $r^\prime$ band.
The legends in the upper right corner of each panel list the $r1\_\Mbh$, $r2\_\Mbh$ and $b\_\Mbh$ for
the study of the variability amplitudes in the $r^\prime$ band, while the texts in the lower right corner
of each panel give the ranges of the restricted parameters. For the radio-loud quasars refer to
Fig. 27 in the Appendix.
\label{sigma_eddratio_mbh}}
\end{figure}

\begin{figure}[]
\plotone{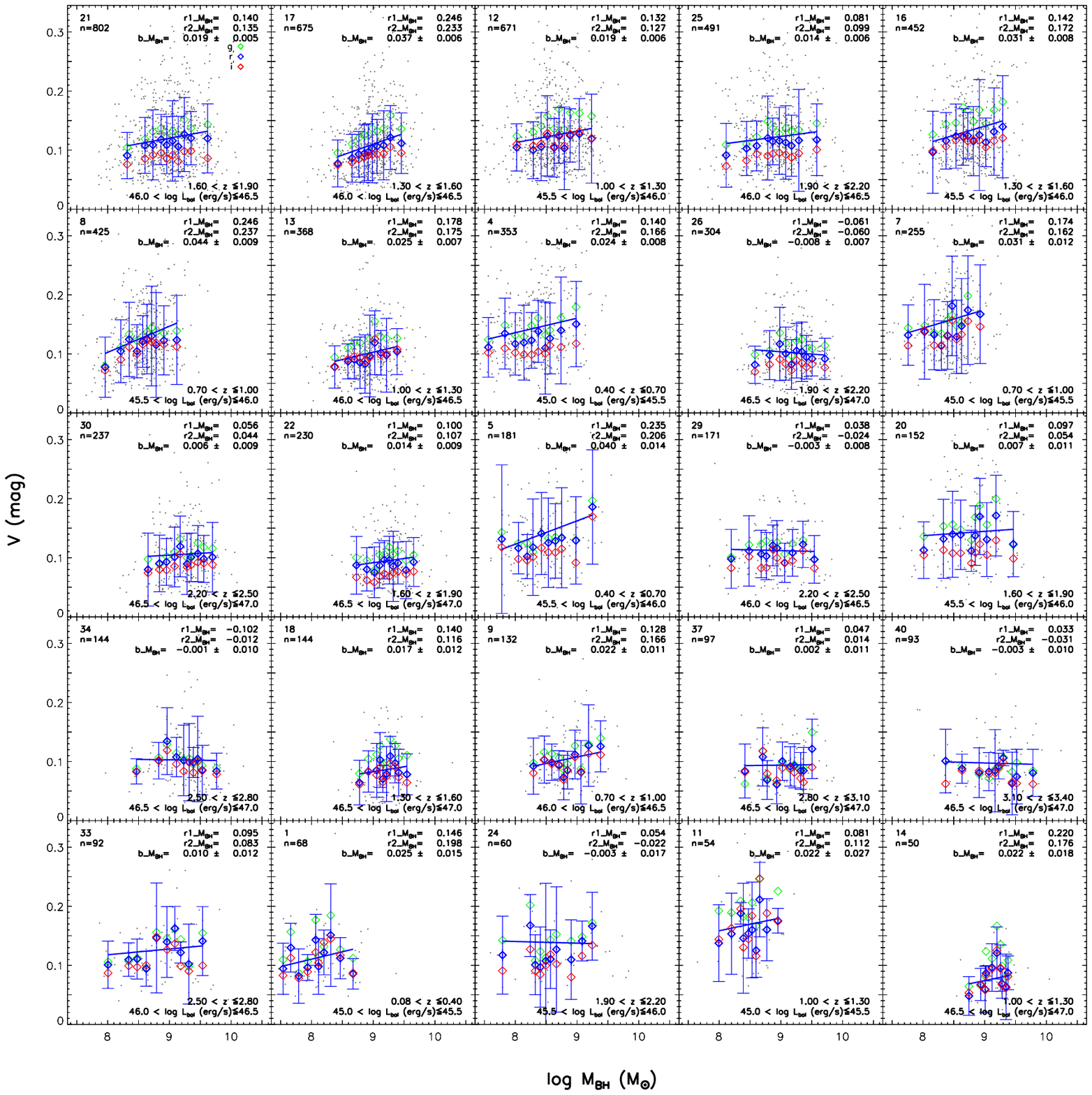}
\caption{Figure similar to Fig.~\ref{sigma_eddratio_mbh}, but for the qualified subsamples obtained
from the $\log{\Lbol}$-$z$ space as illustrated in Fig.~\ref{binlbol}. Refer to Fig. 28 in
the Appendix for the radio-loud quasars.
\label{sigma_lbol_mbh}}
\end{figure}

    The relation between the variability amplitude and $\Mbh$ for the 25 largest qualified data
sets obtained from binning in the $\log{\Redd}$-$z$ space is shown in Fig.~\ref{sigma_eddratio_mbh}, and
that for the subsamples with restricted $\Lbol$ values is shown in Fig.~\ref{sigma_lbol_mbh}. Almost all
the subsamples in the former figure report an apparent inverse dependence of $V$ on $\log{\Mbh}$.
Proceeding from the first panel to the last one in the latter figure, a positive relation is shown in general.
Most subsamples show a significant positive $\log{\Mbh}$ dependence, especially for those
containing more quasars (the first 8 panels). The opposite trend shown in some cases (such as the 3rd panel
in the lowest line) is very weak and there are too few cases to claim deviations from the general trend.

The calculated $r1\_\Mbh$, $p\_\Mbh$, $r2\_\Mbh$ and $b\_\Mbh$ values for each subsample are shown in
Fig.~\ref{zdata_mbh}. In the subsamples obtained from binning in the $\log{\Redd}$-$z$ space,
we can find that for both the radio-quiet and the radio-loud quasars, there is a significant
anti-correlation between them with the median coefficient $r1\_\Mbh$ about -0.203 above 94\% confidence
level and -0.220 at a lower confidence level ($p\_{\Mbh} \sim$ 0.432), respectively.
While for the radio-quiet subsamples with $\Lbol$ limited, more than 71.1 percent show positive
$r1\_\Mbh$ values and its median value is 0.113 ($p\_{\Mbh} \sim$ 0.235); the median value of
0.246 ($p\_{\Mbh} \sim$ 0.386) is shown for the radio-loud quasars, suggesting an insignificant correlation.
Based on the median values of the linear fitting results ($b\_{\Mbh}$ and $a\_{\Mbh}$), we can parameterize
the relation between $V$ and $\Mbh$ using Eq.~(\ref{lamequation}) with $X$ as $\log{\Mbh}$, where
$b\_\Mbh = -0.033 \pm 0.019$ and $a\_\Mbh = 0.394 \pm 0.168$ for the subsamples binned from the
$\log{\Redd}$-$z$ space;
$b\_\Mbh = 0.013 \pm 0.014$ and $a\_\Mbh = -0.002 \pm 0.119$ for the subsamples given by binning in
the $\log{\Lbol}$-$z$ space. The median values of $b\_\Mbh$ and $a\_\Mbh$  are -0.052$\pm$0.052,
0.600$\pm$0.477 in the $\log{\Redd}$-$z$ space and 0.040$\pm$0.035, -0.206$\pm$0.309 in the
$\log{\Lbol}$-$z$ space for the radio-loud quasars. It guarantees the information provided by the
distribution of $r1\_\Mbh$---the relations are insignificant for the radio-loud subsamples.
The median $r1$, $p$, $r2$, $b$, $b_{\rm err}$, $a$ and $a_{\rm err}$ values for the relationships between
$V$ and different quasar parameters are all summarized in Table~\ref{reltotal}.

\section{Discussion\label{discussion}}
\subsection{Comparison with previous work}
\cite{VB04} argued that variability of radio-loud quasars are larger than radio-quiet ones,
while no evidence for a change in the average optical variability amplitude with increasing
radio luminosity is shown in \cite{Bauer09}.
Comparing the median variability amplitude, we do not find larger variability amplitudes for
radio-loud quasars than radio-quiet ones.
However, their intrinsic distributions are different based on the results of KS test.
The variability of radio-loud quasars may be caused by some mechanisms other
than the accretion disk instability, such as the outflow \citep{Giveon99}.

We confirm the anti-correlation between the variability amplitude and
rest-frame wavelength which is mostly studied in the form of
ensemble structure functions \citep{Cutri85, Paltani94, VB04}.
We show that both the radio-quiet quasars and radio-loud quasars
exhibit a bluer-when-brighter chromatism, in good accordance
with previous studies\citep{Wilhite05, Meusinger10, Trevese01,
Trevese02, Schmidt12, Ai11}.
\cite{Honma91} showed that when instabilities occur in the inner disk,
the integrated color will change because of the outward propagating hot wave,
which can account for the bluer-when-brighter behavior. \cite{Sakata11} also
suggested that it may be caused by changes of accretion rate. Besides
this explanation, radiations from different regions of accretion disk or influences of
host galaxy may also induce such a behavior \citep{Shields78, Malkan83, Hawkins93}.
However, as quasar optical luminosities are dominated by their disks rather than their
host galaxies, accretion disk instabilities are more likely than influences
of host galaxies to explain the bluer-when-brighter chromatism.
Trevese \& Vagnetti (2001, 2002) analyzed the dependence of spectral
changes on the brightness of quasars and suggested that the chromatism is at least
consistent with the temperature variation of a single emitting blackbody and can
be explained by bright spots on the disk possibly produced by the accretion disk
instability, which also agrees with \cite{Schmidt12}. It is also possible that
a combination of these physical mechanisms produces the color chromatism.

We find no relation between the variability amplitude and redshift for radio-quiet 
quasars, which seems contrary to the negative or positive correlations 
reported in previous work \citep{Cristiani90, Hook94, Trevese94}. The anti-correlation found
in \cite{Cristiani90} may reflect the anti-correlation between $V$ and $\Lbol$,
as their sample is too small to disentangle the relation. Only
\cite{Trevese94} binned the magnitude limited sample within small luminosity and redshift
ranges and found an anti-correlation, which is ascribed to the dependence of variability
on wavelength. Some previous work \citep{Giveon99, Schmidt12} also found no relation
between $V$ and $z$. After correcting the contribution from emission lines, the color
variability is independent of redshift \citep{Schmidt12}, which agrees with our results.
However, although \cite{Giveon99} also found no relation between $V$ and $z$, their conclusion
is probably not reliable due to the small size and the narrow redshift range of the quasar sample.
Our results do isolate the influences of rest-frame wavelength,
black hole mass and Eddington ratio and suggest that the redshift dependence of radio-quiet
quasar variability may not exist. The possible anti-correlation between $V$ and $z$ for
radio-loud quasars is very weak and not significant; it may need to be investigated with
a larger sample in the future.

By investigating the relationships between the variability amplitude and the intrinsic quasar parameters
in different redshift ranges, both before and after binning in one of the parameters,
we find some results that are different from those in previous work.
In spite of obvious selection effects of the flux-limited sample
within $0<z<0.75$, \cite{Wold07} reported a significant dependence
of variability on black hole mass, approximately 0.2 magnitude
increase of the variability value over 2-3 orders of magnitude in
black hole mass. They used the maximum magnitude difference of
light curves as the variability indicator, which is approximately 4
times larger than our estimated variability value.
Before considering the relations among the intrinsic quasar parameters, the quasars within
$0.08<z\le0.4$ and $0.4<z\le0.7$ in our sample also show a relation between
the quasar variability amplitude and black hole mass.
However, this is not the whole story for the correlation between variability
and $\Mbh$. We do not find any certain relation between the variability amplitudes and 
black hole mass in the subsamples binned in the intrinsic quasar parameters.
Most subsamples in narrow ranges of Eddington ratio exhibit a weak anti-correlation 
while the subsamples in narrow luminosity ranges show no relation or a weak correlation 
between variability values and black hole mass.
We derived the best-fit parameters with the function similar to Eq. (10) in \cite{Bauer09}
for our subsamples in small bins of luminosity.
Although the resulting slope for the largest subsample is 0.067$\pm$0.018,
from statistical results of all qualified radio-quiet samples we do not find the
weak correlation stated in \cite{Bauer09}. The median slope is 0.051$\pm$0.056
compared to 0.13$\pm$0.01 in their work.

\cite{Wilhite08} also found a correlation between the variability amplitude
and black hole mass and an anti-correlation between the variability amplitude and
luminosity independent of black hole mass. The black hole mass range
of our sample ($10^7$ M$_{\odot}$ - $10^{10.8}$ M$_{\odot}$) is larger than that of their sample,
which makes our results statistically more significant. Compared to their
results, the relation between the variability amplitude and luminosity is weaker
in our case. This may be due to the fact that in our sample, many bright (faint) quasars
also show large (small) variability, which further shallows the anti-correlation.
Furthermore, this corresponds well to the anti-correlation found in small
$\Mbh$ bins in \cite{Bauer09}, where the logarithm value of the variability amplitude is
proportional to (-0.205 $\pm$ 0.002) $\log{\Lbol}$.
With our sample, we calculate the slope using the fitting function similar to theirs and
the slope value is -0.263$\pm$0.074. We confirm the anti-correlation between variability and 
$\Redd$ \citep{Wilhite08, Ai10} by excluding the influences of other parameters on the 
variability amplitude.

\subsection{Implications of the standard accretion disk model}
The optical continuum of low-redshift quasars can often be well fitted by the
standard accretion-disk model proposed by \cite{Shakura73}.
Following \cite{Li08}, we investigate the dependencies of the variability
amplitude on rest-frame wavelength and the intrinsic quasar parameters,
assuming that quasar variability is caused by a change of accretion rate,
which is the result of accretion disk instabilities. However, we do not utilize the temperature
distribution shown in their work, as it is actually an approximation to the equatorial temperature
in the outer region of the disk and depends on viscosity. The temperature distribution we used
here describes the radial dependence of the effective temperature derived from the emergent flux
from an accretion disk around a Schwarzchild black hole, which is independent of viscosity, namely,
\begin{equation}
\label{temperature}
T(R)=\left[\frac{3 G M \dot{M}}{8 \pi \sigma R^3} \left(1-\sqrt{R_{\rm in}/{R}}\right)\right]^{1/4},
\end{equation}
where
$R_{\rm in}=3 R_{\rm g}$,
$R_{\rm g}=\frac{2 GM}{c^2}$.
Here, $\sigma$ is the Stefan-Boltzmann constant, $R$ is the radius of the disk,
$M$ is the black hole mass, $\dot{M}$ is the accretion rate.
     After calculating the emitting spectrum at each radius, we integrate it
over the whole disk to get the final spectrum. The emitted spectrum of the disk can be written as:
\begin{equation}
\label{spectrum}
f_\nu = \frac{4\pi h \cos(\beta) \nu^3}{c^2 D^2} \int^{R_{\rm out}}_{R_{\rm in}}
\frac{R\,dR}{e^{h\nu/k T(R)}-1},
\end{equation}
where $R_{\rm out}$ is the outer radius of the disk, $h$ is Plank's constant,
$\beta$ is the inclination of the disk with respect to the line of sight and $D$
is the luminosity distance from the observer to the quasar \citep{Frank02}.
We simply assume the inclination to be 0. It is a reasonable assumption since any inclination other
than 0 will just reduce the normalization of the whole spectra, which does not affect the variability
amplitude. In addition, the inclination is not likely to change significantly.

    To better compare theoretical results with observations, we obtain
$\eta \dot{m}$ values in the model according to the representative $\Redd$ values
in our sample.
\begin{equation}
\dot{M}=\dot{m} \dot{M}_{\rm crit} = \eta \dot{M}_{\rm EDD} \dot{m},
\end{equation}
where $\dot{M}_{\rm crit}=L_{\rm EDD}/c^2$ is the critical accretion rate, and
$\dot{M}_{\rm EDD}=L_{\rm EDD}/(\eta c^2)$ is the Eddington accretion rate.
Based on this equation, $\dot{m}$ in \cite{Li08} is $\eta$ ( $\sim$ 1/16) times
our $\dot{m}$ as their $\dot{m}$ is the ratio of $\dot{M}$ to $\dot{M}_{\rm EDD}$.

\begin{figure}[]
\plotone{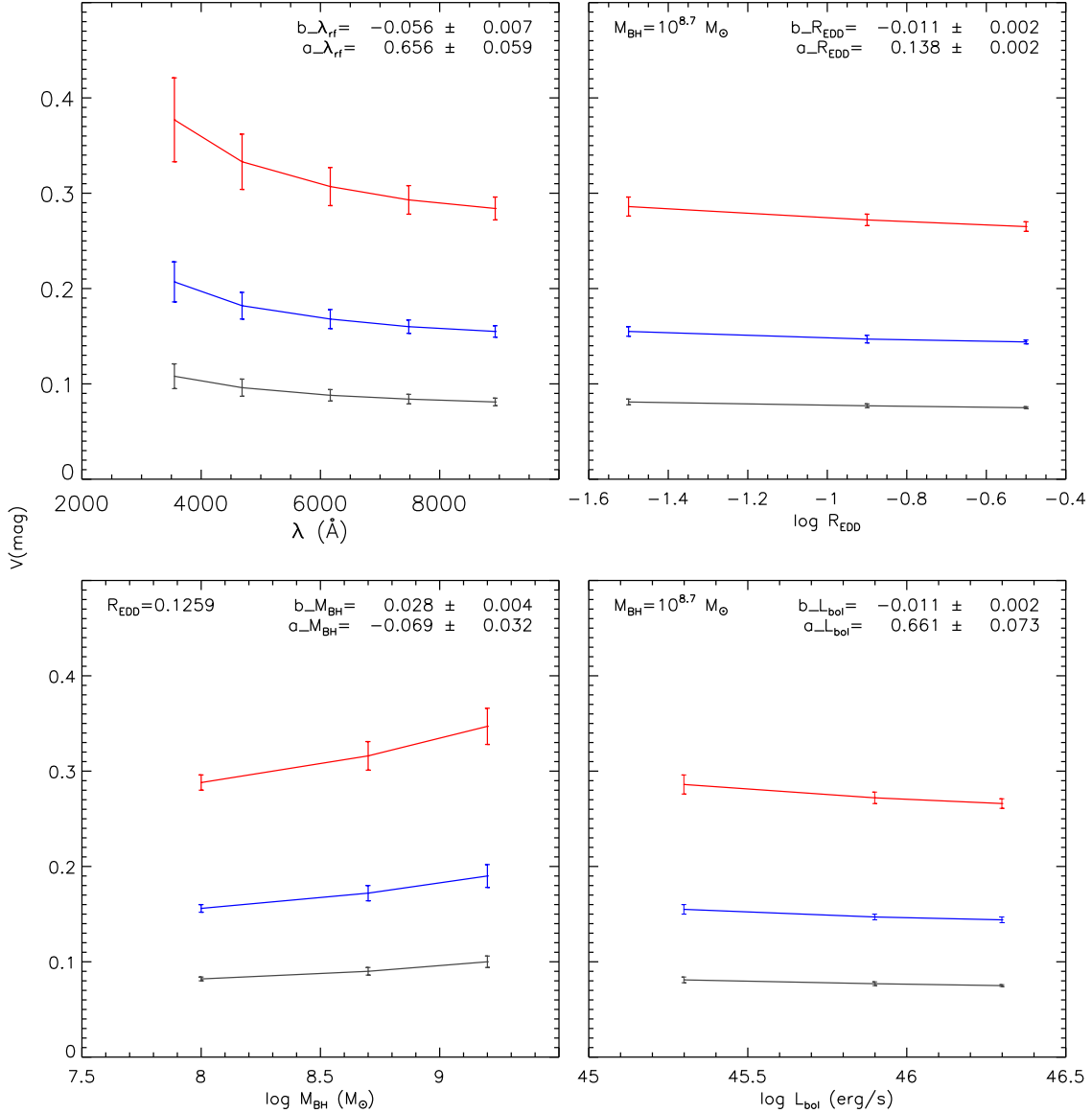}
\caption{ The upper left panel shows the model-predicted variability amplitude as a function
of wavelength, here $M_{\rm BH} = 10^{8.7}$ M$_{\odot}$ and $\Redd = 0.1259$; the upper right panel
shows the variability amplitude in $r^\prime$ band as a function of $\Redd$.
The lower left panel shows the variability amplitude in the $r^\prime$ band as a function of
$\Mbh$; the lower right panel shows the variability amplitude in the $r^\prime$ band
as a function of $\Lbol$. The red, blue and black lines refer to the cases
with different changes of accretion rate, namely 0.4$\dot{m}$, 0.2$\dot{m}$ and
0.1$\dot{m}$. The obtained slope $b$ values for the simulated data during the case with the
change of accretion rate at 0.2$\dot{m}$ are shown in the upper right corner of each panel.
\label{flux_param}}
\end{figure}

      We calculate spectra for different black hole masses and accretion rates.
$\Mbh$ equals to $10^8$, $10^{8.7}$ and $10^9$ M$_\odot$ with an observational
uncertainty of 0.4 dex \citep{Vestergaard06}, while $\Redd$ equals to 
0.0316, 0.1259 and 0.3162 with an uncertainty of 0.3 dex \citep{Kollmeier06}.
After convolving with the SDSS filter response function, we convert the flux to the SDSS
AB magnitude $A(\Mbh,\dot{m})$.
Assuming the change of $\dot{m}$ is $x \dot{m}$, we adopt 
$V=A(\Mbh,\dot{m})-A(\Mbh, (1+x) \dot{m})$ as variability
indicators. Results for the three $x$ values (0.4, 0.2, 0.1) are shown as red, blue and
black lines respectively in Fig.~\ref{flux_param}.
At a specific $x$, the variability amplitudes along with their errors are calculated as below.
A set of 100 normally-distributed random $\Mbh$ around $10^{8.7}$ M$_{\odot}$ and a
set of 100 normally-distributed random $\Redd$ around 0.1259 are carried 
out within their observational uncertainties.
Based on the set of $\Mbh$ and the set of $\Redd$, we estimate variability
amplitudes in each filter band to see its dependence on rest-frame wavelength.
The mean value of the set of estimations and 3 times the standard deviation
in each filter ($u^\prime$, $g^\prime$, $r^\prime$, $i^\prime$ and $z^\prime$) are adopted as
the final variability and its uncertainty respectively, as shown in the upper left panel of
Fig.~\ref{flux_param}.
The variability amplitude declines with the increasing of rest-frame wavelength under the
three changes of accretion rate.
In the upper right panel, an anti-correlation between $V$ in the $r^\prime$ band
and $\Redd$ is displayed, where $V$ and the error in $V$ are respectively the mean value and 3 times the
standard deviation of a set of variability amplitudes estimated from 100 random realizations
of the model\footnote{$\Mbh$ in the model
is generated from a normal distribution with a mean value of $10^{8.7}$ M$_\odot$ and an observational
uncertainty of 0.4 dex, $\dot{m}$ is one of the three different $\dot{m}$ and $x \dot{m}$ refers
to a change of the corresponding $\dot{m}$.}.
As luminosity is only related to $\Redd$ if $\Mbh$ is fixed, an inverse relation with $\Lbol$
is naturally expected. We do see this trend in the lower right panel of Fig.~\ref{flux_param}.
For a set of 100 normally-distributed random $\Redd$ values around 0.1259 within its observational
uncertainty, we obtained 3 sets of $V$ values, each set with respect to one of the three $\Mbh$ values.
The mean value and 3 times the dispersion of $V$ values in each set is adopted as the final $V$ and
error in $V$ for the $\Mbh$ value.
As displayed in the lower left panel, there is an increase of optical variability with the increase
of black hole mass.
Comparisons between our theoretic results and observed relationships are carried out
in each panel. We find that the variability amplitude caused by a change of accretion rate around
0.2$\dot{m}$ agrees well with the majority of variability amplitudes in observations.
Moreover, besides the relation of variability with rest-frame wavelength, the model
also reproduces the tendencies of correlations between the variability amplitude and
luminosity, as well as accretion rate.
We fit the simulated data points as we do with the real data in the $x=0.2$ case, which are
shown in the upper right of each panel. All the anti-correlations are flatter than the results
from observations.
The predicted correlation seems to be inconsistent with the weak anti-correlation
between the variability amplitude and $\Mbh$ for the subsamples within small ranges of $\Redd$ as
we stated in \textsection~\ref{param}. It suggests that other physical mechanisms may still
need to be considered in modifying the simple accretion disk model.

\section{Conclusion\label{conclusion}}
A sample of 7658 quasars drawn from SDSS-DR7 is used to
investigate the relationship between the variability amplitude and rest-frame
wavelength, redshift and the intrinsic quasar parameters, i.e., black hole
mass, bolometric luminosity and Eddington ratio. We analyzed
the radio-quiet and radio-loud quasar variability properties with the
sample. Because we have better quality data points for each quasar and a large
sample, the variability indicator is calculated for each quasar from its
long term light curve and the dependence of different parameters
are disentangled by binning the parameter space into small slices.
Our main conclusions are the following:

1. There is an anti-correlation between the variability amplitude and
rest-frame wavelength after excluding the influences of redshift
and the intrinsic parameters on variability amplitude.
The quasar variability amplitude of radio-quiet quasars seems not dependent on redshift,
while the possible anti-correlation between that of radio-loud quasars and redshift 
still needs be confirmed with a larger sample.

2. Although the median variability amplitude of the radio-loud quasars is similar
compared to that of radio-quiet quasars, their intrinsic distributions are different.
It implies that there may be some mechanisms other than the accretion disk instability
behind the variability of radio-loud quasars.

3. There is a bluer-when-brighter color chromatism for both radio-quiet quasars and
radio-loud ones. Some mechanisms, including instabilities occurring in the inner disk,
the influence of the host galaxy, bright spots or a combination of them, may explain
the chromatism.

4. The variability amplitude is significantly anti-correlated with $\Redd$
and $\Lbol$ in the subsamples without binning the quasar
parameters. There is no certain correlation between the variability amplitude and $\Mbh$
in all these subsamples. After considering the relationships among the intrinsic quasar
parameters, we find that small data sets still exhibit an anti-correlation
between the variability amplitude and $\Redd$, and between the variability amplitude and
$\Lbol$ with a lower significance level. Moreover, an anti-correlation
of the variability amplitude with $\Mbh$ begins to emerge in bins of $\Redd$, but
a positive correlation with $\Mbh$ is shown in bins of $\Lbol$.

5. Based on the Shakura-Sunyaev model, assuming the change of accretion rate as the origin of
quasar variability, we reproduce the relationship of the variability amplitude
with rest-frame wavelength with good agreement, and the tendencies of correlations between
variability and $\Redd$ as well as $\Lbol$.
The predicted correlation between the variability amplitude and $\Mbh$ seems not to correspond well
to the case inferred from observations, namely an anti-correlation with $\Mbh$
in small bins of $\Redd$. It implies that the change of accretion rate is important
for producing the observed optical variability but other physical mechanisms still
need to be considered in modifying the simple accretion disk model for quasars.

We thank the anonymous referee for suggestion that have significantly improved
the quality of this paper.
We also thank Fukun Liu for helpful discussions on the relationship between
variability and black hole mass. We acknowledge Tie Liu and Song Huang for helpful
discussions on the statistical test of distributions of variability amplitudes.
Thank Stephen Justham, Zhao-Yu Li and Awat Rahimi for reading the manuscript and providing various
suggestions that have greatly improved the draft. This work was supported by an
NSFC grant (No. 11033001) and the 973 program (No. 2007CB8154505) in China.
Funding for the SDSS and SDSS-II has been provided by the Alfred P.
Sloan Foundation, the Participating Institutions, the National Science
Foundation, the U.S. Department of Energy, the National Aeronau- tics
and Space Administration, the Japanese Monbukagakusho, the Max Planck
Society, and the Higher Education Funding Council for England. The
SDSS Web Site is http://www.sdss.org/.
The SDSS is managed by the Astrophysical Research Consortium for the
Participating Institutions. The Participating Institutions are the
American Museum of Natural History, Astrophysical Institute Potsdam,
University of Basel, University of Cambridge, Case Western Reserve
University, University of Chicago, Drexel University, Fermilab, the
Institute for Advanced Study, the Japan Participation Group, Johns
Hopkins University, the Joint Institute for Nuclear Astrophysics, the
Kavli Institute for Particle Astrophysics and Cosmology, the Korean
Scientist Group, the Chinese Academy of Sciences (LAMOST), Los Alamos
National Laboratory, the Max-Planck-Institute for Astronomy (MPIA),
the Max- Planck-Institute for Astrophysics (MPA), New Mexico State
University, Ohio State University, University of Pittsburgh,
University of Portsmouth, Princeton University, the United States
Naval Observatory, and the University of Washington.

\clearpage

\begin{deluxetable}{ccccccccccccccccccccccccccccccccc}
\tabletypesize{\scriptsize}
\setlength{\tabcolsep}{0.05in}
\tablecolumns{10} \tablewidth{0pc}
\tablecaption{The bin details of each parameter\label{parambin}}
\tablehead{
\colhead{Parameter}   &
\colhead{Minimum} & \colhead{Maximum} & \colhead{Bin size} & \colhead{N$_{\rm bins}$}\\
\\
\colhead{(1)} & \colhead{(2)} & \colhead{(3)} &
\colhead{(4)} & \colhead{(4)}
}
\startdata
$z$          &  0.08& 4.9  & 0.3   & 16 \\
$\log{\Mbh}$ (M$_\odot)$ &  7.5 & 10.5    &0.5 &  6    \\
$\log{\Lbol}$ (erg$/$s)&  44.5 & 48   & 0.5 &  7  \\
$\log{\Redd}$ &  -2.0 & 0.5    & 0.5  & 5 \\
\enddata
\tablecomments{
Col. (1) Parameter names.
Col. (2) Minimum value of the binned parameter in Col. (1).
Col. (3) Maximum value of the binned parameter in Col. (1).
Col. (4) Bin interval of the binned parameter in Col. (1). Note the first bin size of $z$ is 0.32.
Col. (5) Number of bins.}
\end{deluxetable}

\begin{deluxetable}{ccccccccccccccccccccccccccccccccc}
\tabletypesize{\scriptsize}
\setlength{\tabcolsep}{0.05in}
\tablecolumns{10} \tablewidth{0pc}
\tablecaption{Correlation between variability and each parameter\label{reltotal}}
\tablehead{
\colhead{Radio loudness} &
\colhead{Parameter}   &
\multicolumn{3}{c}{Restricted Parameters} &
\multicolumn{2}{c}{Spearman relation} &
\colhead{Pearson relation} &
\multicolumn{2}{c}{Linear fit} \\
\colhead{} & \colhead{} &
\colhead{$x1$} & \colhead{$x2$} &\colhead{$x3$} &
\colhead{$r1$} & \colhead{$p$}
& \colhead{$r2$} &
\colhead{$b$} & \colhead{$a$}       \\
\\
\colhead{(1)} & \colhead{(2)} & \colhead{(3)} &
\colhead{(4)} & \colhead{(5)} & \colhead{(6)} &
\colhead{(7)} & \colhead{(8)} & \colhead{(9)} &
\colhead{(10)}
}
\startdata
    radio-quiet    &      $\log{\lamrf}$   &      $\Redd$   &            $z$   &         $\Mbh$  &    -0.229 &      0.019  &    -0.233  &    -0.137$\pm$ 0.052  &     0.579 $\pm$ 0.180    \\
    radio-quiet    &      $\log{z}$        &      $\Redd$   &         $\Mbh$   &       $\lamrf$  &    -0.081 &      0.189  &    -0.090  &    -0.049$\pm$ 0.044  &     0.131 $\pm$ 0.009    \\
    radio-quiet    &      $\log{\Lbol}$    &          $z$   &        $\Redd$   &       $\lamrf$  &    -0.269 &      0.015  &    -0.260  &    -0.044$\pm$ 0.020  &     2.128 $\pm$ 0.913    \\
    radio-quiet    &      $\log{\Lbol}$    &          $z$   &         $\Mbh$   &       $\lamrf$  &    -0.343 &      0.003  &    -0.335  &    -0.063$\pm$ 0.020  &     3.032 $\pm$ 0.926    \\
    radio-quiet    &      $\log{\Redd}$    &          $z$   &        $\Lbol$   &       $\lamrf$  &    -0.167 &      0.137  &    -0.147  &    -0.020$\pm$ 0.014  &     0.091 $\pm$ 0.011    \\
    radio-quiet    &      $\log{\Redd}$    &          $z$   &         $\Mbh$   &       $\lamrf$  &    -0.303 &      0.013  &    -0.299  &    -0.052$\pm$ 0.019  &     0.075 $\pm$ 0.013    \\
    radio-quiet    &      $\log{\Mbh}$     &          $z$   &        $\Lbol$   &       $\lamrf$  &     0.113 &      0.235  &     0.104  &     0.013$\pm$ 0.014  &    -0.002 $\pm$ 0.119    \\
    radio-quiet    &      $\log{\Mbh}$     &          $z$   &        $\Redd$   &       $\lamrf$  &    -0.203 &      0.057  &    -0.201  &    -0.033$\pm$ 0.019  &     0.394 $\pm$ 0.168    \\
     radio-loud    &      $\log{\lamrf}$   &      $\Redd$   &            $z$   &         $\Mbh$  &    -0.187 &      0.218  &    -0.224  &    -0.225$\pm$ 0.143  &     0.950 $\pm$ 0.494    \\
     radio-loud    &      $\log{z}$        &      $\Redd$   &         $\Mbh$   &       $\lamrf$  &    -0.153 &      0.306  &    -0.062  &    -0.034$\pm$ 0.116  &     0.148 $\pm$ 0.028    \\
     radio-loud    &      $\log{\Lbol}$    &          $z$   &        $\Redd$   &       $\lamrf$  &    -0.266 &      0.297  &    -0.269  &    -0.058$\pm$ 0.054  &     2.794 $\pm$ 2.456    \\
     radio-loud    &      $\log{\Lbol}$    &          $z$   &         $\Mbh$   &       $\lamrf$  &    -0.384 &      0.164  &    -0.321  &    -0.071$\pm$ 0.055  &     3.373 $\pm$ 2.536    \\
     radio-loud    &      $\log{\Redd}$    &          $z$   &        $\Lbol$   &       $\lamrf$  &    -0.304 &      0.300  &    -0.274  &    -0.045$\pm$ 0.032  &     0.079 $\pm$ 0.035    \\
     radio-loud    &      $\log{\Redd}$    &          $z$   &         $\Mbh$   &       $\lamrf$  &    -0.369 &      0.188  &    -0.291  &    -0.069$\pm$ 0.047  &     0.056 $\pm$ 0.055    \\
     radio-loud    &      $\log{\Mbh}$     &          $z$   &        $\Lbol$   &       $\lamrf$  &     0.246 &      0.386  &     0.212  &     0.040$\pm$ 0.035  &    -0.206 $\pm$ 0.309    \\
     radio-loud    &      $\log{\Mbh}$     &          $z$   &        $\Redd$   &       $\lamrf$  &    -0.220 &      0.432  &    -0.180  &    -0.052$\pm$ 0.052  &     0.600 $\pm$ 0.477    \\
\enddata
\tablecomments{
Col. (1) The samples, namely the radio-quiet (radio-loud) subsample contains quasars with radio loudness smaller (larger) than 10.
Col. (2) The parameter on which the dependence of variability is to be considered.
Cols. (3-5) Parameters which are restricted in small ranges to exclude their influences on variability, denoted by $x$1, $x$2 and $x$3.
Cols. (6-7) $r1$ is the median value of Spearman rank correlation coefficients in all the qualified small data sets, which are obtained from binning
in the restricted parameters, while $p$ is the median value of the significance of its deviation from 0.
Col. (8) $r2$ is the median value of Pearson Product Correlation Coefficient, indicating the strength of linear correlations.
Cols. (9-10) $b$ is the median slope of the linear fitting for the dependence of the variability amplitude on the parameter denoted in Col.(2) and $a$ is the median y axis intercept of the linear fit.}
\end{deluxetable}

\clearpage

\clearpage

\clearpage

\clearpage

\appendix
\section{Appendix}
For the radio-loud quasars, we use the same mechanism to investigate correlations between $V$ and
quasar parameters. The detailed relations are shown here.
\begin{figure}[]
\plotone{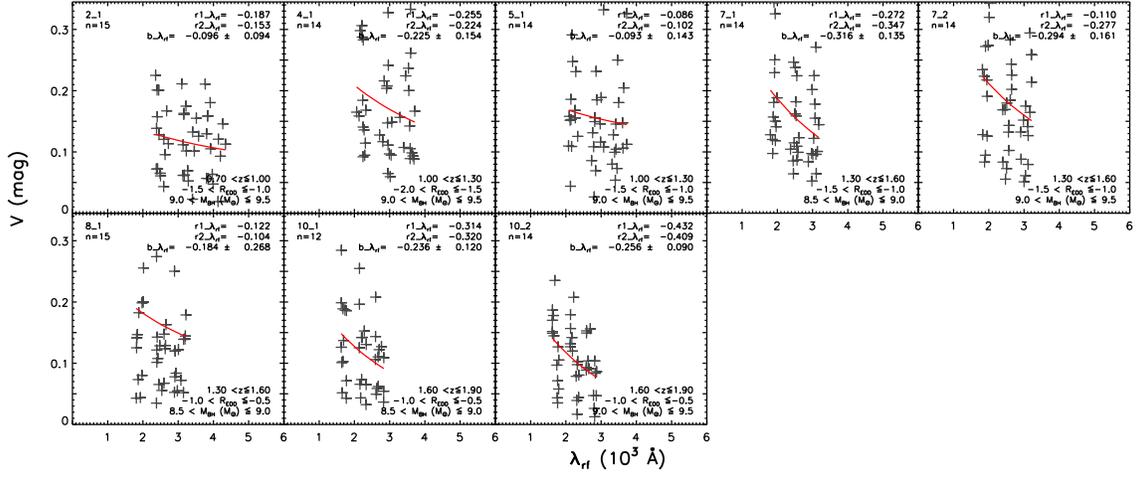}
\caption{Dependence of $V$ on $\lamrf$ in the $g^\prime$, $r^\prime$ and $i^\prime$ band for all the 8
radio-loud sub-subsamples within small ranges of $\Redd$, $z$ and $\Mbh$. In each panel, the red curve
refers to the polynomial fitting with the function similar to Eq. (4), where X is $\lamrf$.
$r1\_\lamrf$, $r2\_\lamrf$ and $b\_\lamrf$ are shown in the upper right corner and ranges of
the restricted parameters are also shown in the lower right corner. Refer to Fig.~\ref{sigma_lamdarq}
for the radio-quiet quasars.
\label{sigma_lamdarl}}
\end{figure}

\begin{figure}[]
\plotone{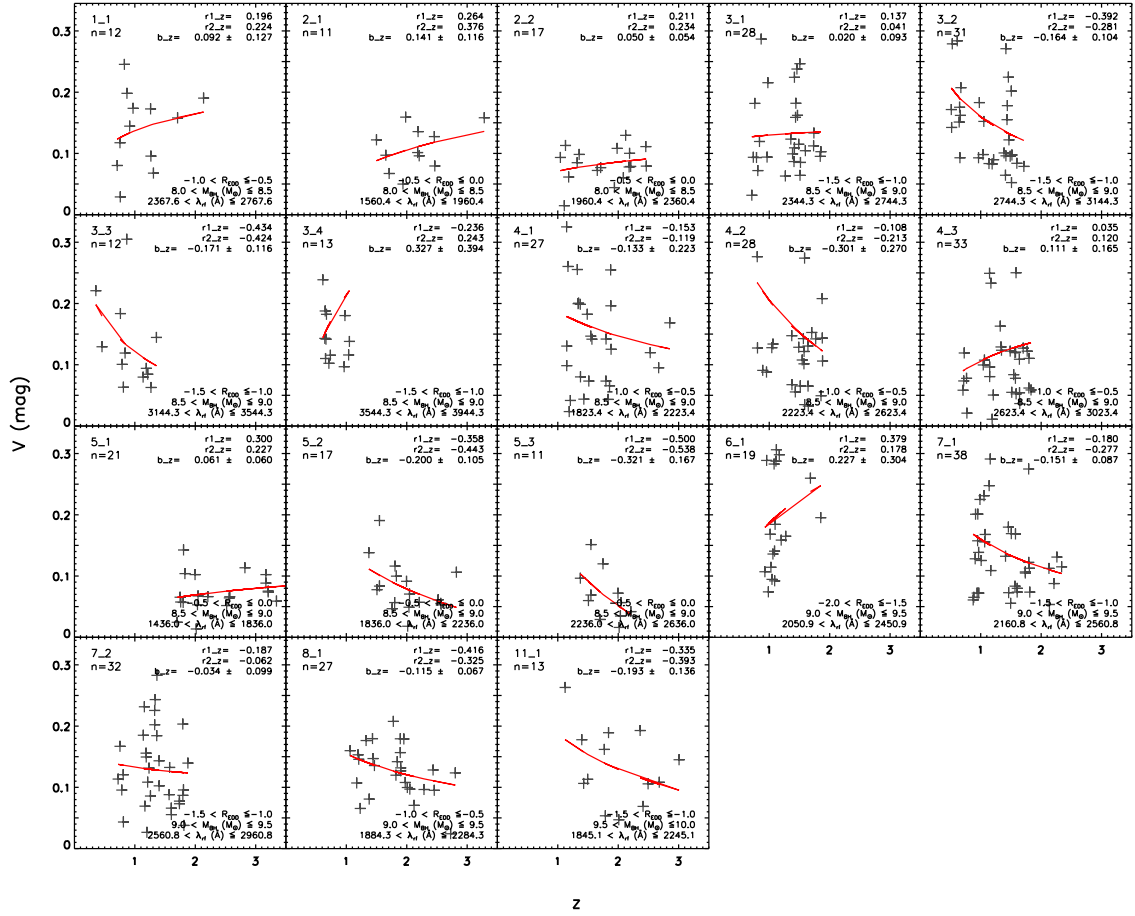}
\caption{Dependence of $V$ on $z$ in the $g^\prime$, $r^\prime$ and $i^\prime$ band and all
the 18 radio-loud sub-subsamples within small ranges of $\Redd$, $\Mbh$ and $\lambda_{\rm rf}$.
$r1\_z$, $r2\_z$ and ranges of the restricted parameters are shown in the upper right
corner of each panel. Refer to Fig.~\ref{sigma_zrq} for the radio-quiet quasars.
\label{sigma_zrl}}
\end{figure}

\begin{figure}[]
\plotone{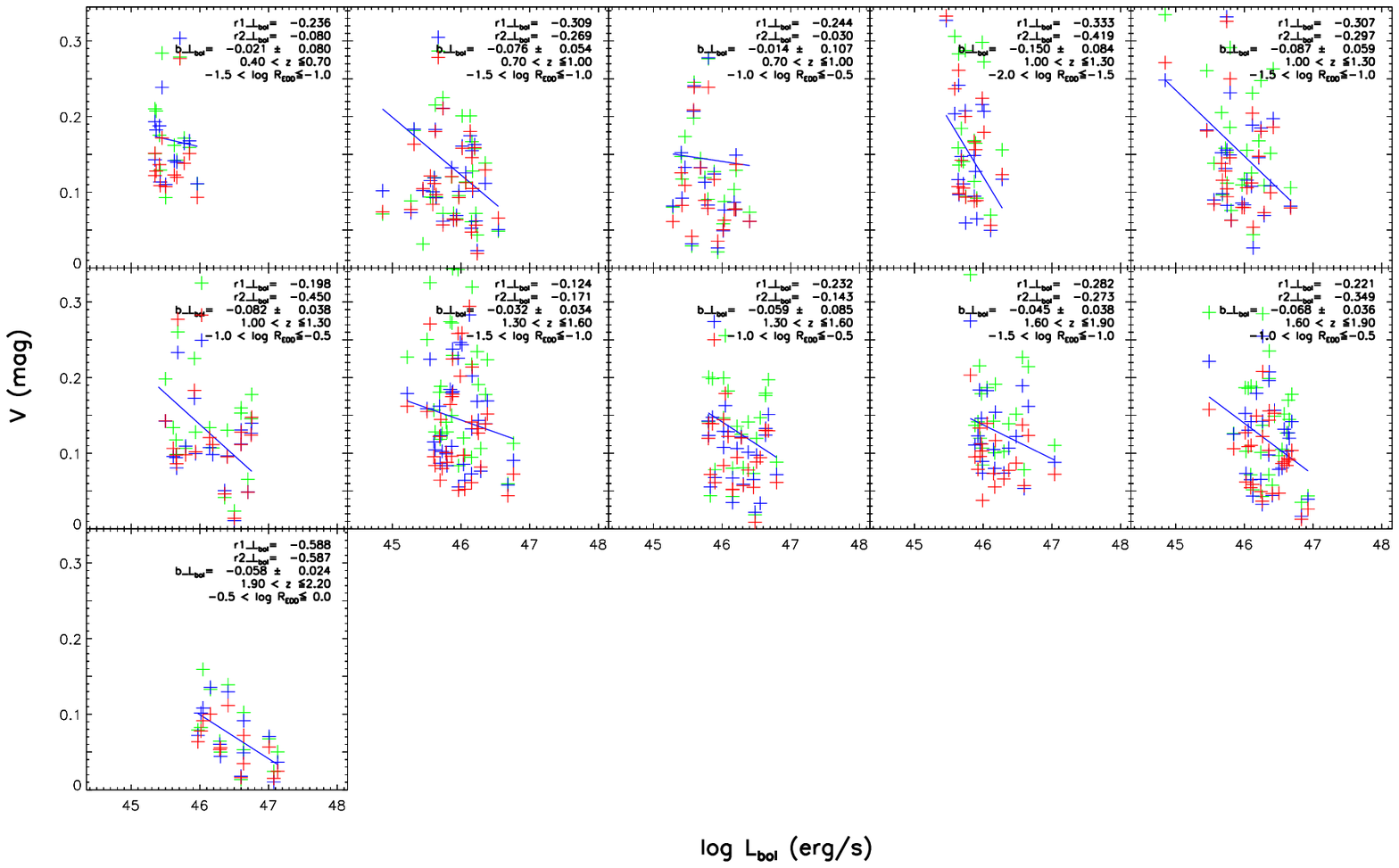}
\caption{Dependence of $V$ in the $g^\prime$, $r^\prime$ and $i^\prime$ filter band on
$\Lbol$ in the qualified subsamples obtained from the $\log{\Redd}$-$z$ space,
represented by green, blue and red plus signs respectively.
The blue line refer to the linear fit for all the data in the $r^\prime$ band.
The legends in the upper right corner of each panel refer to the $r1\_\Lbol$, $r2\_\Lbol$
and $b\_\Lbol$ for the study of the variability amplitudes in the $r^\prime$ band and ranges
of the restricted parameters. Refer to Fig.~\ref{sigma_eddratio_lbol} for the radio-quiet quasars.
\label{sigma_eddratio_lbolrl}}
\end{figure}

\begin{figure}[]
\plotone{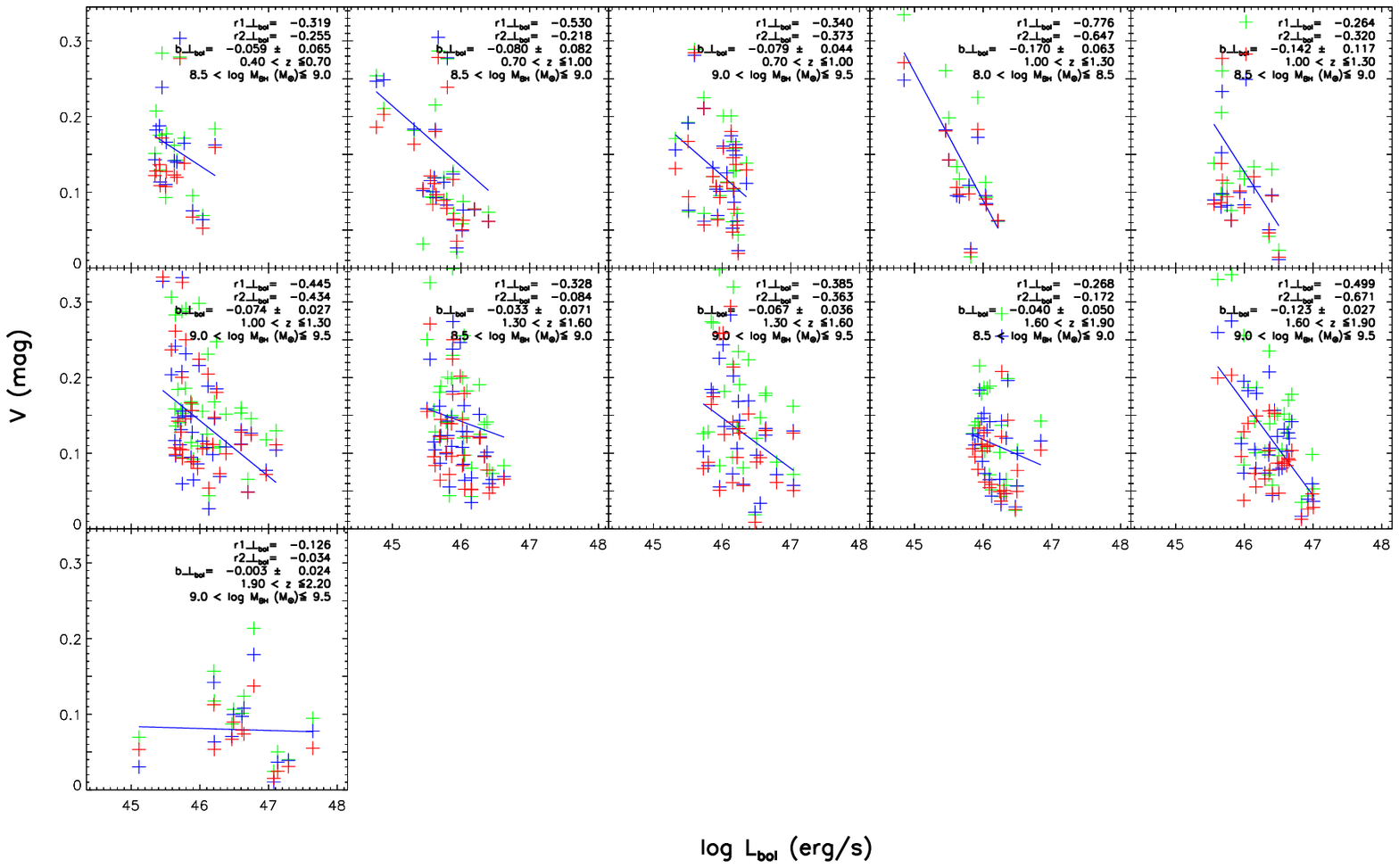}
\caption{Figure similar to Fig. 23, but for the qualified subsamples obtained in the $\log{\Mbh}$-$z$ space.
Refer to Fig.~\ref{sigma_mbh_lbol} for the radio-quiet quasars.
\label{sigma_mbh_lbolrl}}
\end{figure}

\begin{figure}[]
\plotone{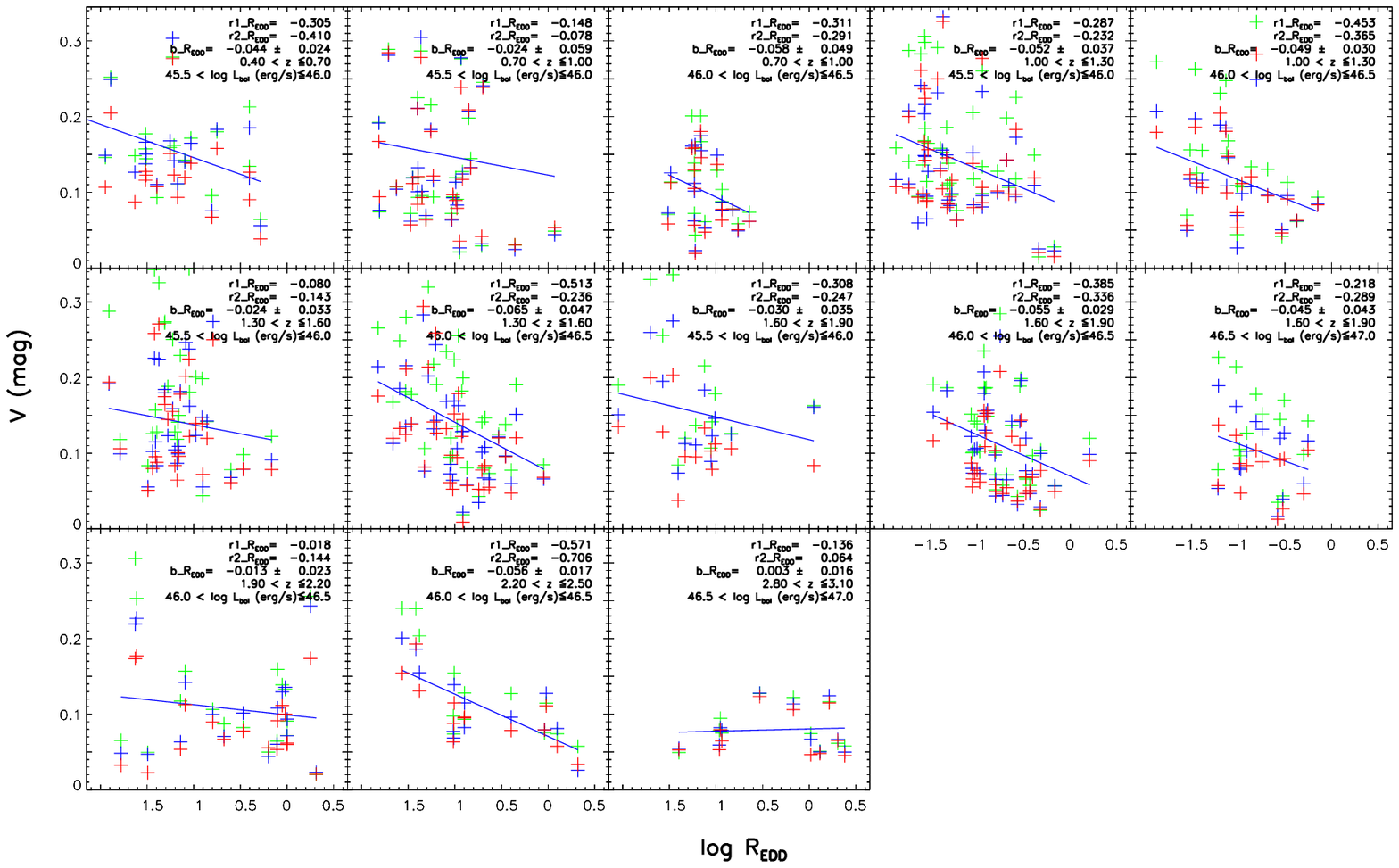}
\caption{Dependence of $V$ in the $g^\prime$, $r^\prime$ and $i^\prime$ filter band on
$\Redd$ in the qualified subsamples obtained from the $\log{\Lbol}$-$z$ space,
represented by green, blue and red plus signs respectively.
The blue lines represent the linear fits for all the data in the $r^\prime$ band.
The legends on the upper right of each panel list $r1\_\Redd$, $r2\_\Redd$ and $b\_\Redd$
for the study of the variability amplitudes in the $r^\prime$ band and ranges of the restricted
parameters. Refer to Fig.~\ref{sigma_lbol_eddratio} for the radio-quiet quasars.
\label{sigma_lbol_eddratiorl}}
\end{figure}

\begin{figure}[]
\plotone{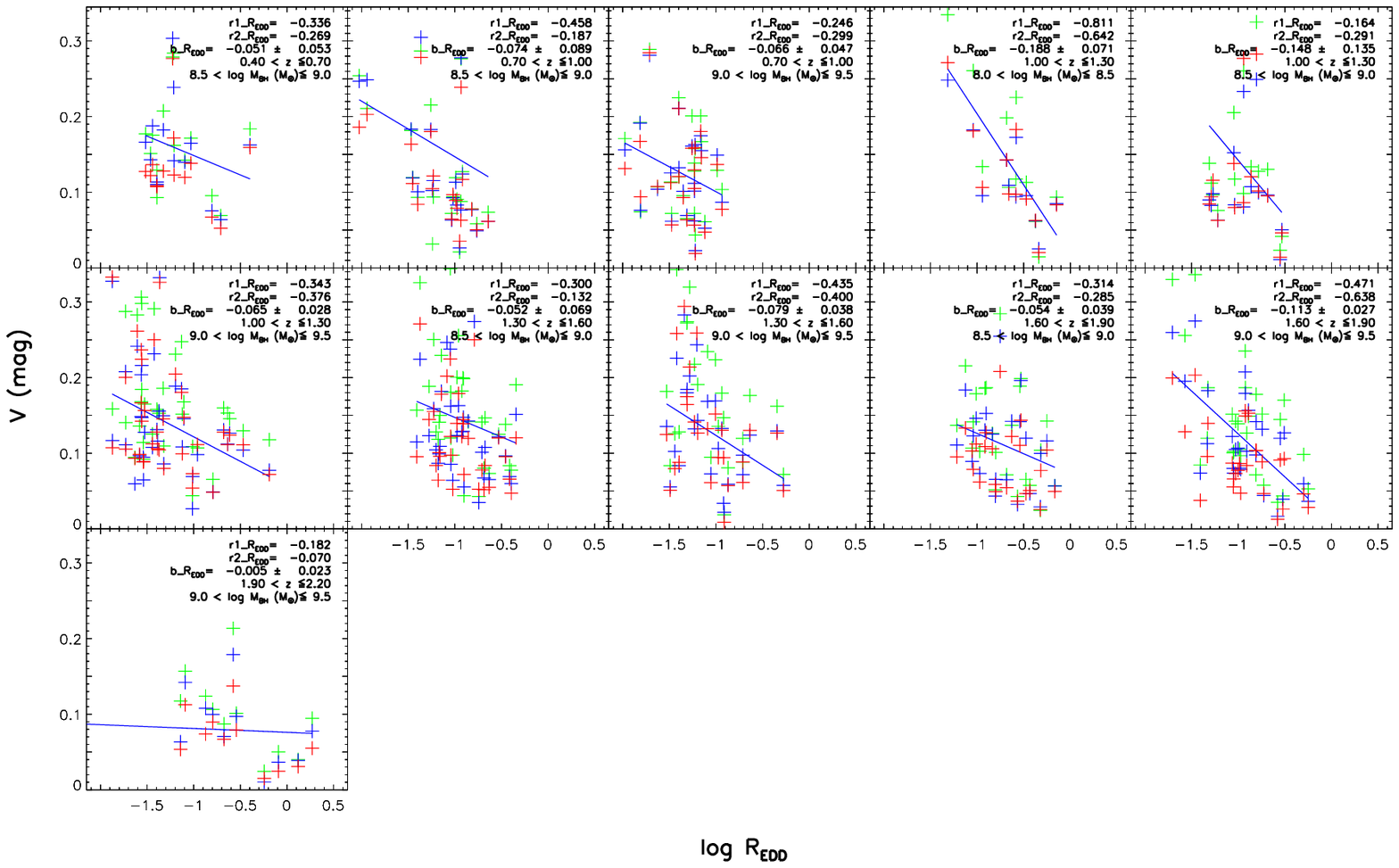}
\caption{Figure similar to Fig. 25, but for the qualified subsamples obtained
in the $\log{\Mbh}$-$z$ space. Refer to Fig.~\ref{sigma_mbh_eddratio} for the radio-quiet quasars.
\label{sigma_mbh_eddratiorl}}
\end{figure}
\clearpage

\begin{figure}[]
\plotone{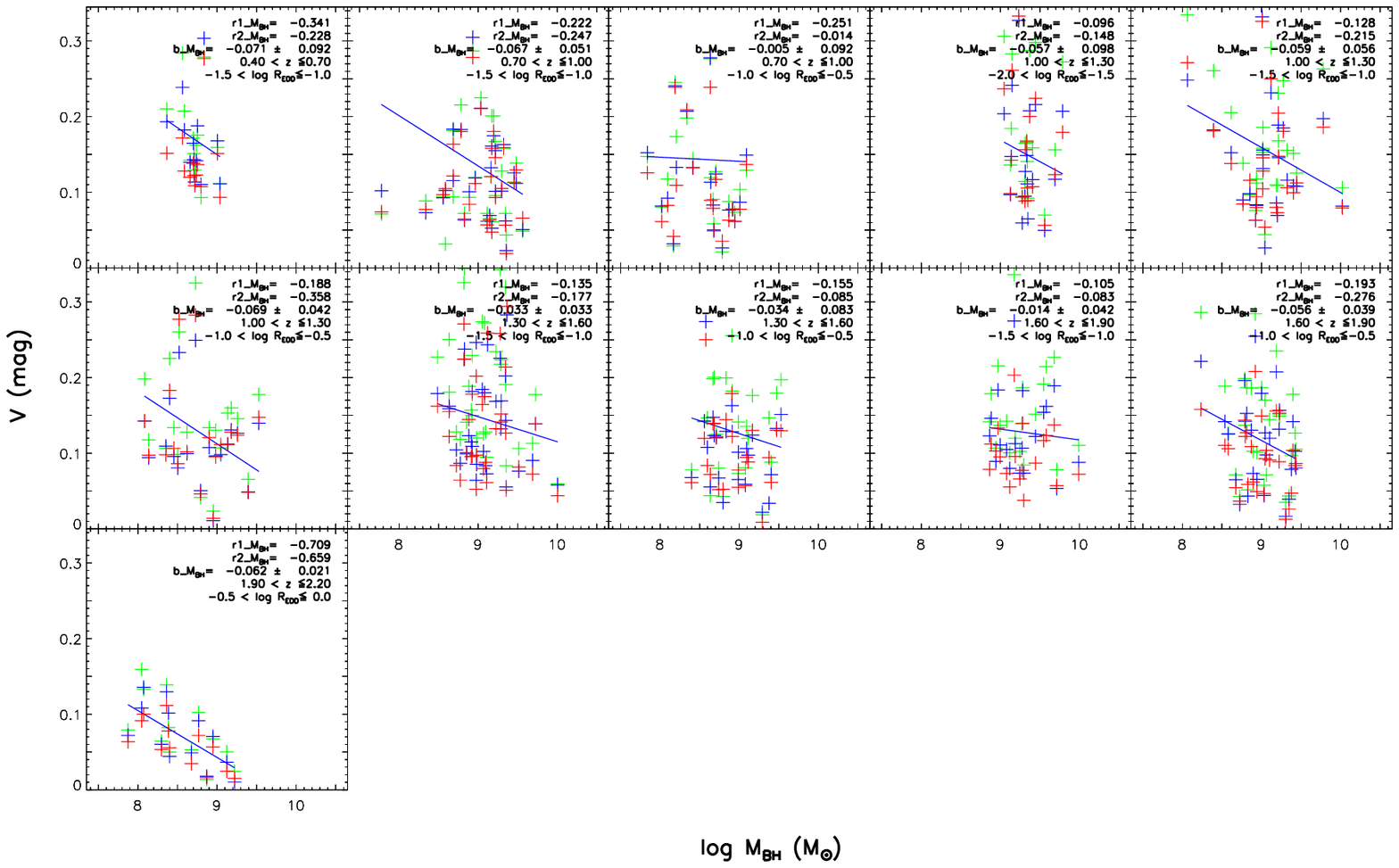}
\caption{Dependence of $V$ in the $g^\prime$, $r^\prime$ and $i^\prime$ filter band on
$\Mbh$ in the qualified subsamples obtained from the $\log{\Redd}$-$z$ space,
represented by green, blue and red plus signs respectively.
The blue lines show the linear fits for all the data in the $r^\prime$ band.
The legends on the upper right of each panel list the $r1\_\Mbh$, $r2\_\Mbh$ and $b\_\Mbh$
for the study of variability amplitude in the $r^\prime$ band and ranges of the restricted
parameters. Refer to Fig.~\ref{sigma_eddratio_mbh} for the radio-quiet quasars.
\label{sigma_eddratio_mbhrl}}
\end{figure}

\begin{figure}[]
\plotone{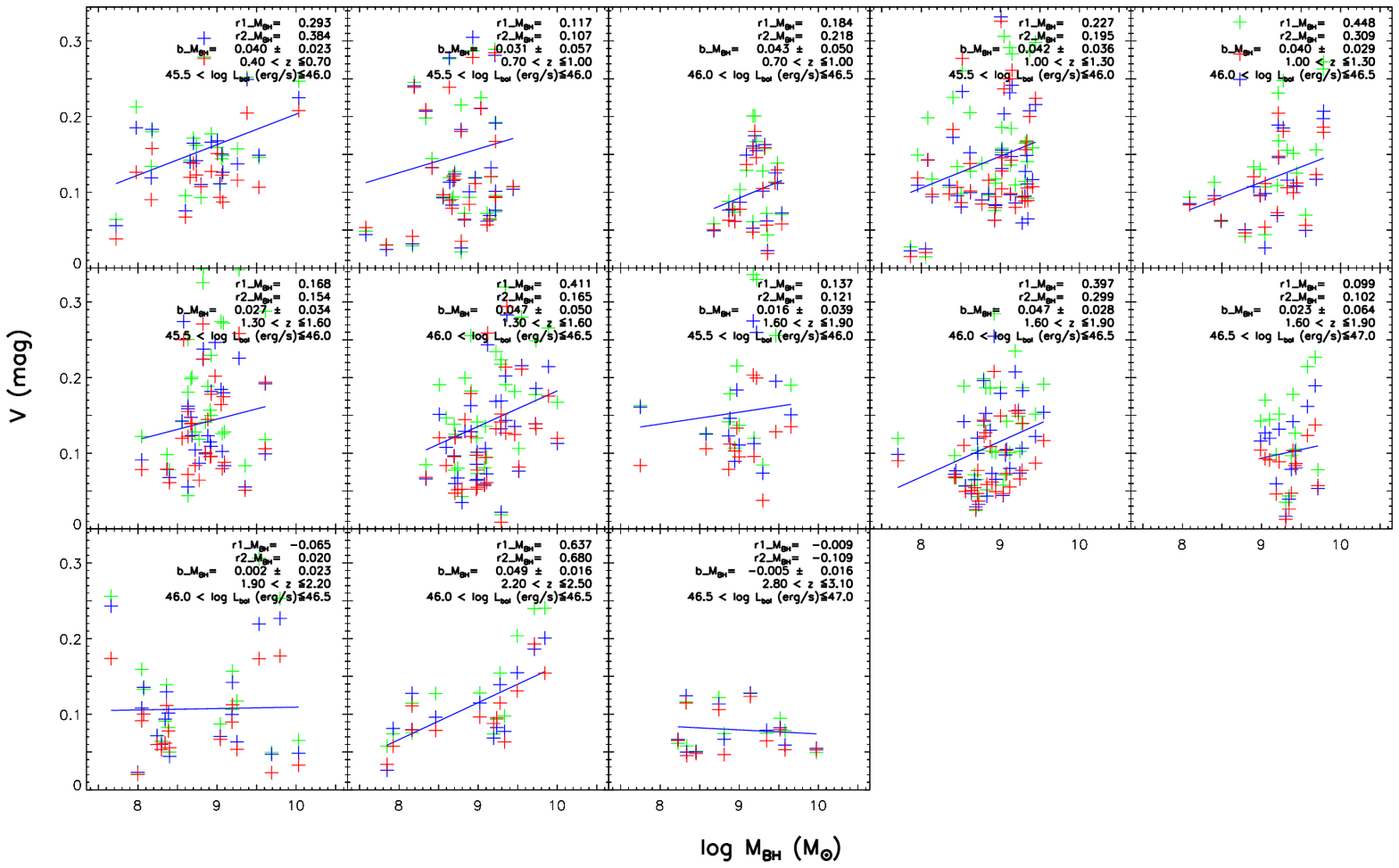}
\caption{Figure similar to Fig. 27, but for the qualified subsamples obtained from the $\log{\Lbol}$-$z$ space.
Refer to Fig.~\ref{sigma_lbol_mbh} for the radio-quiet quasars.
\label{sigma_lbol_mbhrl}}
\end{figure}


\begin{thebibliography}{}
\bibitem[Ai et al.(2010)]{Ai10} Ai, Y.~L., Yuan, W., Zhou,
H.~Y., et al.\ 2010, \apjl, 716, L31


\bibitem[Ai et al.(2011)]{Ai11} Ai, Y.~L., Yuan, W., Zhou,
H.~Y., Wang, T.~G., \& Zhang, S.~H.\ 2011, \apj, 727, 31


\bibitem[Aretxaga et al.(1997)]{Aretxaga97} Aretxaga, I., Cid
Fernandes, R., \& Terlevich, R.~J.\ 1997, \mnras, 286, 271


\bibitem[Bauer et al.(2009)]{Bauer09} Bauer, A., Baltay, C.,
Coppi, P., et al.\ 2009, \apj, 696, 1241


\bibitem[Becker et al.(1995)]{Becker95} Becker, R.~H., White,
R.~L., \& Helfand, D.~J.\ 1995, \apj, 450, 559


\bibitem[Bonoli et al.(1979)]{Bonoli79} Bonoli, F., Braccesi, A., Federici, L., Zitelli, V., \& Formiggini, L.\ 1979, \aaps, 35, 391


\bibitem[Butler \& Bloom(2011)]{Butler11}Butler, N.~R., \& Bloom, J.~S.\ 2011, \aj, 141, 93


\bibitem[Cid Fernandes et al.(2000)]{Cid00} Cid Fernandes,
R., Sodr{\'e}, L., Jr., \& Vieira da Silva, L., Jr.\ 2000, \apj, 544, 123


\bibitem[Courvoisier et
al.(1996)]{Courvoisier96} Courvoisier, T.~J.-L., Paltani, S., \& Walter, R.\ 1996, \aap, 308, L17


\bibitem[Cristiani et
al.(1996)]{Cristiani96} Cristiani, S., Trentini, S., La Franca, F., et al.\ 1996, \aap, 306, 395


\bibitem[Cristiani et al.(1990)]{Cristiani90} Cristiani, S., Vio,
R., \& Andreani, P.\ 1990, \aj, 100, 56


\bibitem[Cutri et al.(1985)]{Cutri85} Cutri, R.~M., Wisniewski,
W.~Z., Rieke, G.~H., \& Lebofsky, M.~J.\ 1985, \apj, 296, 423


\bibitem[Czerny et al.(2008)]{Czerny08} Czerny, B.,
Siemiginowska, A., Janiuk, A., \& Gupta, A.~C.\ 2008, \mnras, 386, 1557


\bibitem[di Clemente et al.(1996)]{Di96} di Clemente, A.,
Giallongo, E., Natali, G., Trevese, D.,
\& Vagnetti, F.\ 1996, \apj, 463, 466


\bibitem[Frank et al.(2002)]{Frank02} Frank, J., King, A.,
\& Raine, D.~J.\ 2002, Accretion Power in Astrophysics, by Juhan Frank and Andrew King and Derek Raine, pp.~398.~ISBN 0521620538.~Cambridge, UK: Cambridge University Press, February 2002.,


\bibitem[Fukugita et al.(1996)]{Fukugita96} Fukugita, M.,
Ichikawa, T., Gunn, J.~E., et al.\ 1996, \aj, 111, 1748


\bibitem[Giallongo et al.(1991)]{Giallongo91} Giallongo, E.,
Trevese, D., \& Vagnetti, F.\ 1991, \apj, 377, 345


\bibitem[Giveon et al.(1999)]{Giveon99} Giveon, U., Maoz, D.,
Kaspi, S., Netzer, H., \& Smith, P.~S.\ 1999, \mnras, 306, 637


\bibitem[Hawkins(1993)]{Hawkins93} Hawkins, M.~R.~S.\ 1993, \nat,
366, 242


\bibitem[Honma et al.(1991)]{Honma91} Honma, F., Kato, S.,
\& Matsumoto, R.\ 1991, \pasj, 43, 147


\bibitem[Hook et al.(1994)]{Hook94} Hook, I.~M., McMahon,
R.~G., Boyle, B.~J., \& Irwin, M.~J.\ 1994, \mnras, 268, 305


\bibitem[Ivezi{\'c} et al.(2007)]{Ivezic07} Ivezi{\'c}, {\v Z}.,
Smith, J.~A., Miknaitis, G., et al.\ 2007, \aj, 134, 973


\bibitem[Kato et al.(1996)]{Kato96} Kato, S., Abramowicz,
M.~A., \& Chen, X.\ 1996, \pasj, 48, 67


\bibitem[Kawaguchi et al.(1998)]{Kawaguchi98} Kawaguchi, T.,
Mineshige, S., Umemura, M., \& Turner, E.~L.\ 1998, \apj, 504, 671


\bibitem[Kollmeier et al.(2006)]{Kollmeier06} Kollmeier, J.~A.,
Onken, C.~A., Kochanek, C.~S., et al.\ 2006, \apj, 648, 128


\bibitem[Li
\& Cao(2008)]{Li08} Li, S.-L., \& Cao, X.\ 2008, \mnras, 387, L41


\bibitem[MacLeod et al.(2010)]{MacLeod10} MacLeod, C.~L.,
Ivezi{\'c}, {\v Z}., Kochanek, C.~S., et al.\ 2010, \apj, 721, 1014


\bibitem[Malkan(1983)]{Malkan83} Malkan, M.~A.\ 1983, \apj, 268,
582


\bibitem[Manmoto et al.(1996)]{Manmoto96} Manmoto, T., Takeuchi,
M., Mineshige, S., Matsumoto, R., \& Negoro, H.\ 1996, \apjl, 464, L135


\bibitem[Meusinger et al.(2010)]{Meusinger10} Meusinger, H., Hinze,
A., \& de, H.~A.\ 2010, VizieR Online Data Catalog, 352, 59037


\bibitem[Netzer
\& Sheffer(1983)]{Netzer83} Netzer, H., \& Sheffer, Y.\ 1983, \mnras, 203, 935


\bibitem[Paltani
\& Courvoisier(1997)]{Paltani97} Paltani, S., \& Courvoisier, T.~J.-L.\ 1997, \aap, 323, 717


\bibitem[Paltani
\& Courvoisier(1994)]{Paltani94} Paltani, S., \& Courvoisier, T.~J.-L.\ 1994, \aap, 291, 74


\bibitem[Rees(1984)]{Rees84} Rees, M.~J.\ 1984, \araa, 22, 471


\bibitem[Sakata et al.(2011)]{Sakata11} Sakata, Y., Morokuma,
T., Minezaki, T., et al.\ 2011, \apj, 731, 50


\bibitem[Schmidt et al.(2010)]{Schmidt10} Schmidt, K.~B., Rix,
H.-W., Jester, S., et al.\ 2010, IAU Symposium, 267, 265


\bibitem[Schmidt et al.(2012)]{Schmidt12} Schmidt, K.~B., Rix,
H.-W., Shields, J.~C., et al.\ 2012, \apj, 744, 147


\bibitem[Schneider et al.(2010)]{Schneider10} Schneider, D.~P.,
Richards, G.~T., Hall, P.~B., et al.\ 2010, \aj, 139, 2360


\bibitem[Sesar et al.(2007)]{Sesar07} Sesar, B., Ivezi{\'c},
{\v Z}., Lupton, R.~H., et al.\ 2007, \aj, 134, 2236


\bibitem[Shakura
\& Sunyaev(1973)]{Shakura73} Shakura, N.~I., \& Sunyaev, R.~A.\ 1973, \aap, 24, 337


\bibitem[Shen
\& Kelly(2010)]{Shen10} Shen, Y., \& Kelly, B.~C.\ 2010, \apj, 713, 41


\bibitem[Shen et al.(2011)]{Shen11} Shen, Y., Richards, G.~T.,
Strauss, M.~A., et al.\ 2011, \apjs, 194, 45


\bibitem[Shields(1978)]{Shields78} Shields, G.~A.\ 1978, \nat,
272, 706


\bibitem[Terlevich et al.(1992)]{Terlevich92} Terlevich, R.,
Tenorio-Tagle, G., Franco, J., \& Melnick, J.\ 1992, \mnras, 255, 713


\bibitem[Torricelli-Ciamponi et
al.(2000)]{Torricelli00} Torricelli-Ciamponi, G., Foellmi, C., Courvoisier, T.~J.-L., \& Paltani, S.\ 2000, \aap, 358, 57


\bibitem[Tr{\`e}vese
\& Vagnetti(2002)]{Trevese02} Tr{\`e}vese, D., \& Vagnetti, F.\ 2002, \apj, 564, 624


\bibitem[Tr{\`e}vese
\& Vagnetti(2001)]{Trevese01} Tr{\`e}vese, D., \& Vagnetti, F.\ 2001, \memsai, 72, 33


\bibitem[Trevese et al.(1994)]{Trevese94} Trevese, D., Kron,
R.~G., Majewski, S.~R., Bershady, M.~A.,
\& Koo, D.~C.\ 1994, \apj, 433, 494


\bibitem[Vanden Berk et al.(2004)]{VB04} Vanden Berk, D.~E.,
Wilhite, B.~C., Kron, R.~G., et al.\ 2004, \apj, 601, 692


\bibitem[Vestergaard
\& Wilkes(2006)]{Vestergaard06} Vestergaard, M., \& Wilkes, B.~J.\ 2001, \apjs, 134, 1


\bibitem[White et al.(1997)]{White97} White, R.~L., Becker,
R.~H., Helfand, D.~J.,
\& Gregg, M.~D.\ 1997, VizieR Online Data Catalog, 8048, 0


\bibitem[Wilhite et al.(2008)]{Wilhite08} Wilhite, B.~C.,
Brunner, R.~J., Grier, C.~J., Schneider, D.~P.,
\& vanden Berk, D.~E.\ 2008, \mnras, 383, 1232


\bibitem[Wilhite et al.(2005)]{Wilhite05} Wilhite, B.~C., Vanden
Berk, D.~E., Kron, R.~G., et al.\ 2005, \apj, 633, 638


\bibitem[Wold et al.(2007)]{Wold07} Wold, M., Brotherton,
M.~S., \& Shang, Z.\ 2007, \mnras, 375, 989

\end{thebibliography}
\end{document}